\documentstyle[12pt]{article}

\catcode`@=11
\def\un#1{\relax\ifmmode\@@underline#1\else
        $\@@underline{\hbox{#1}}$\relax\fi}
\catcode`@=12

\let\du=\d                      

\def\a{\alpha}
\def\b{\beta}
\def\c{\chi}
\def\d{\delta}
\def\e{\epsilon}
\def\f{\phi}
\def\g{\gamma}
\def\h{\eta}
\def\i{\iota}
\def\j{\psi}

\def\m{\mu}
\def\n{\nu}
\def\o{\omega}
\def\p{\pi}
\def\q{\theta}
\def\r{\rho}
\def\s{\sigma}
\def\t{\tau}

\def\x{\xi}

\def\F{\Phi}
\def\G{\Gamma}
\def\J{\Psi}
\def\L{\Lambda}

\def\S{\Sigma}
\def\U{\Upsilon}
\def\X{\Xi}

\def\ve{\varepsilon}
\def\vf{\varphi}

\def\ca{{\cal A}}
\def\cb{{\cal B}}

\def\ch{{\cal H}}

\def\bo{{\raise-.5ex\hbox{\large$\Box$}}}               
\def\pa{\partial}                                       
\def\pr{\prod}                                          
\def\TH{{\raise.2ex\hbox{$\displaystyle \bigodot$}\mskip-4.7mu \llap H \;}}
\def\face{{\raise.2ex\hbox{$\displaystyle \bigodot$}\mskip-2.2mu \llap {$\ddot
        \smile$}}}                                      
\def\dg{\sp\dagger}                                     

\def\sp#1{{}^{#1}}                              
\def\Bar#1{\overline{#1}}                       
\def\ket#1{\left| #1\right\rangle}              
\def\VEV#1{\left\langle #1\right\rangle}        
\def\abs#1{\left| #1\right|}                    
\def\leftrightarrowfill{$\mathsurround=0pt \mathord\leftarrow \mkern-6mu
        \cleaders\hbox{$\mkern-2mu \mathord- \mkern-2mu$}\hfill
        \mkern-6mu \mathord\rightarrow$}
\def\dvec#1{\vbox{\ialign{##\crcr
        \leftrightarrowfill\crcr\noalign{\kern-1pt\nointerlineskip}
        $\hfil\displaystyle{#1}\hfil$\crcr}}}           
\def\dt#1{{\buildrel {\hbox{\LARGE .}} \over {#1}}}     

\def\frac#1#2{{\textstyle{#1\over\vphantom2\smash{\raise.20ex
        \hbox{$\scriptstyle{#2}$}}}}}                   
\def\ha{\frac12}                                        
\def\sfrac#1#2{{\vphantom1\smash{\lower.5ex\hbox{\small$#1$}}\over
        \vphantom1\smash{\raise.4ex\hbox{\small$#2$}}}} 
\def\bfrac#1#2{{\vphantom1\smash{\lower.5ex\hbox{$#1$}}\over
        \vphantom1\smash{\raise.3ex\hbox{$#2$}}}}       
\def\afrac#1#2{{\vphantom1\smash{\lower.5ex\hbox{$#1$}}\over#2}}    

\def\[{\lfloor{\hskip 0.35pt}\!\!\!\lceil}
\def\]{\rfloor{\hskip 0.35pt}\!\!\!\rceil}

\def\du#1#2{_{#1}{}^{#2}}
\def\ud#1#2{^{#1}{}_{#2}}

\def\fracm#1#2{\hbox{\large{${\frac{{#1}}{{#2}}}$}}}
\def\half{{\fracm12}}
\def\ha{\half}
\def\tr{{\rm tr}}

\def\ula{{\underline a}}

\def\un{\underline}
\def\fracmm#1#2{{{#1}\over{#2}}}

\def\low#1{{\raise -3pt\hbox{${\hskip 0.75pt}\!_{#1}$}}}

\def\Dot#1{\buildrel{_{_{\hskip 0.01in}\bullet}}\over{#1}}
\def\dt#1{\Dot{#1}}

\def\ua{\uparrow}
\def\da{\downarrow}
\def\sdot{\!\cdot\!}

\newskip\humongous \humongous=0pt plus 1000pt minus 1000pt
\def\caja{\mathsurround=0pt}
\def\eqalign#1{\,\vcenter{\openup2\jot \caja
        \ialign{\strut \hfil$\displaystyle{##}$&$
        \displaystyle{{}##}$\hfil\crcr#1\crcr}}\,}
\newif\ifdtup

\def\ref#1{$\sp{#1)}$}

\def\pl#1#2#3{Phys.~Lett.~{\bf {#1}B} (19{#2}) #3}
\def\np#1#2#3{Nucl.~Phys.~{\bf B{#1}} (19{#2}) #3}
\def\prl#1#2#3{Phys.~Rev.~Lett.~{\bf #1} (19{#2}) #3}
\def\pr#1#2#3{Phys.~Rev.~{\bf D{#1}} (19{#2}) #3}
\def\cqg#1#2#3{Class.~and Quantum Grav.~{\bf {#1}} (19{#2}) #3}

\def\mpl#1#2#3{Mod.~Phys.~Lett.~{\bf A{#1}} (19{#2}) #3}

\parskip=\medskipamount                 
\thicklines                         
\begin{document}


\thispagestyle{empty}               

\def\border{                                            
        \setlength{\unitlength}{1mm}
        \newcount\xco
        \newcount\yco
        \xco=-24
        \yco=12
        \begin{picture}(140,0)
        \put(-20,11){\tiny Institut f\"ur Theoretische Physik Universit\"at
Hannover~~ Institut f\"ur Theoretische Physik Universit\"at Hannover~~
Institut f\"ur Theoretische Physik Hannover}
        \put(-20,-241.5){\tiny Institut f\"ur Theoretische Physik Universit\"at
Hannover~~ Institut f\"ur Theoretische Physik Universit\"at Hannover~~
Institut f\"ur Theoretische Physik Hannover}
        \end{picture}
        \par\vskip-8mm}

\def\headpic{                                           
        \indent
        \setlength{\unitlength}{.8mm}
        \thinlines
        \par
        \begin{picture}(29,16)
        \put(75,16){\line(1,0){4}}
        \put(80,16){\line(1,0){4}}
        \put(85,16){\line(1,0){4}}
        \put(92,16){\line(1,0){4}}

        \put(85,0){\line(1,0){4}}
        \put(89,8){\line(1,0){3}}
        \put(92,0){\line(1,0){4}}

        \put(85,0){\line(0,1){16}}
        \put(96,0){\line(0,1){16}}
        \put(79,0){\line(0,1){16}}
        \put(80,0){\line(0,1){16}}
        \put(89,0){\line(0,1){16}}
        \put(92,0){\line(0,1){16}}
        \put(79,16){\oval(8,32)[bl]}
        \put(80,16){\oval(8,32)[br]}

        \end{picture}
        \par\vskip-6.5mm
        \thicklines}

\border\headpic {\hbox to\hsize{
\vbox{\noindent DESY 93--191 \\
ITP--UH--24/93 \hfill December 1993 \\
hep-th/9312150 \hfill corrected August 1994}}}

\vskip1.3cm
\begin{center}

{\Large\bf Twisting the $N{=}2$ String}
\footnote{Supported in part by the `Deutsche Forschungsgemeinschaft'}\\
\vglue.2in

Sergei V. Ketov \footnote{
On leave of absence from:
High Current Electronics Institute of the Russian Academy of Sciences,
Siberian Branch, Akademichesky~4, Tomsk 634055, Russia}
$\;$,\hspace{.5cm}
Olaf Lechtenfeld

{\it Institut f\"ur Theoretische Physik, Universit\"at Hannover}\\
{\it Appelstra\ss{}e 2, 30167 Hannover, Germany}\\
{\sl ketov, lechtenf @itp.uni-hannover.de}

and

Andrew J. Parkes

{\it Department of Artificial Intelligence}\\
{\it 80 South Bridge, Edinburgh EH1 9HN, U.K.}\\
\end{center}
\vglue.2in
\begin{center}
{\Large\bf Abstract}
\end{center}

The most general homogeneous monodromy conditions in $N{=}2$ string theory
are classified in terms of the conjugacy classes of the global symmetry group
$U(1,1)\otimes{\bf Z}_2$.
For classes which generate a discrete subgroup $\G$,
the corresponding target space backgrounds ${\bf C}^{1,1}/\G$
include half spaces, complex orbifolds and tori.
We propose a generalization of the intercept formula to matrix-valued twists,
but find massless physical states only for $\Gamma{=}{\bf 1}$ (untwisted) and
$\Gamma{=}{\bf Z}_2$ (\`a la Mathur and Mukhi), as well as for $\Gamma$ being
a parabolic element of $U(1,1)$.
In particular, the sixteen ${\bf Z}_2$-twisted sectors of the $N{=}2$ string
are investigated, and the corresponding ground states are identified via
bosonization and BRST cohomology.
We find enough room for an extended multiplet of `spacetime' supersymmetry,
with the number of supersymmetries being dependent on global `spacetime'
topology.
However, world-sheet locality for the chiral vertex operators
does not permit interactions among all massless `spacetime' fermions.
\vglue.2in

\newpage
\hfuzz=10pt
\section{Introduction}

Since the discovery of $N{=}2$ supersymmetric critical strings
in 1976~\cite{aba,aba2} their
status was undergoing several fundamental changes. Initially constructed
in {\it two} spacetime dimensions, they were lately recognized as the strings
naturally living in a {\it four}-dimensional spacetime of apparently
non-physical $(4,0)$ or $(2,2)$ signature~\cite{ft,ali}. The interacting
$N{=}2$
string theory was shown to be closely related to {\it self-dual}
four-dimensional field theories, and it was even conjectured to be the
`master theory' for all integrable models~\cite{ov}.
Recently, some arguments were presented that the $N{=}2$ string
should also be able to support {\it target}
(`spacetime') self-dual supersymmetry, by relating it to the $N{=}4$
fermionic string theory~\cite{siegel}. The appropriate framework for
extended self-dual supersymmetry and supergravity in $2{+}2$~dimensions was
developed in ref.~\cite{kng}. Very recently, $N{=}2$ strings reappeared in
a quite different context of {\it universal} string theory including the
conventional $N{=}0$ and $N{=}1$ strings as particular vacua~\cite{bv}.

In spite of all these amazing developments,
the underlying symmetries and the physical spectrum of the naively `simple'
(compared to the others) $N{=}2$ string theory remain to be poorly understood.
It is known from the calculation of the $N{=}2$ string partition function
on a torus that there exists just a single massless spacetime boson
in the spectrum of the {\it untwisted} $N{=}2$ string moving
in the flat $(2{+}2)$-dimensional background~${\bf R}^{2,2}$~\cite{ov}.
The study of the related $N{=}2$ BRST cohomology with at least some
continuous spacetime momenta was recently initiated in ref.~\cite{gr}. In
addition, there are elements of the BRST cohomology corresponding to {\it
discrete} states at vanishing momentum, $k{=}0$. They can be most easily
identified and investigated when using a compactified background
${\bf T}^{2,2}$ instead of ${\bf R}^{2,2}$~\cite{gs,li}, but discrete
states are not going to be the subject of this paper. In order to derive the
{\it complete} spectrum of the critical $N{=}2$ string in a flat background,
and
to address, in particular, the issue of `spacetime' fermionic physical states
(with continuous momenta), one must investigate the most general monodromy
conditions for the $N{=}2$ string and their associated BRST cohomologies.

We are going to investigate in this paper {\it all\/} possible homogeneous
twistings of the $N{=}2$ string, and find those leading to consistent
solutions. We distinguish a hierarchy of three different types of twists,
namely, with increasing generality:
(i) those flipping only signs in the (bosonic) monodromies,
(ii) those creating arbitrary phases around world-sheet cycles, and
(iii) those mixing different string coordinates, leading to non-compact
monodromies in a diagonal or Jordan normal basis. Clearly, the first type
comprises the simplest (and, we believe, the most important) generalizations
of the naive Neveu-Schwarz-- and Ramond--type boundary conditions, and we are
going to analyze them first. As for the second type of monodromies, the role
of the {\it spectral flow\/} present in the $N{=}2$ superconformal algebra has
to be understood. Finally, the third monodromy type implies rather unusual
topologies of the string target space, which may explain why, to our knowledge,
 it has not been explored in the past. We are particularly interested in
backgrounds which allow {\it massless\/} physical states, in order to analyze
the underlying effective field theory. This does not yet mean that the other
backgrounds are inconsistent. Restricting ourselves to twisted $N{=}2$ strings
 with massless ground states, we find only three possibilities:
(i) the untwisted string as studied e.g. by Ooguri and Vafa \cite{ov},
(ii) implementing the previously known Mathur-Mukhi twist \cite{mm}, and
(iii) a peculiar background corresponding to parabolic elements of $U(1,1)$.
We proceed with the BRST cohomology analysis only in the first two cases,
in order to establish restrictions on interactions of the Mathur-Mukhi
twisted states.

The paper is organized as follows. In sect.~2, we start with a
general discussion of boundary conditions in string theory, formulate the
framework for our subsequent investigation of general  $N{=}2$ string
boundary conditions, and set up our notation. In sect.~3 we recapitulate all
the known local and global, continuous and discrete symmetries of the
Brink-Schwarz $N{=}2$ string action, and list four types of ${\bf Z}_2$ twist
symmetries. This allows us to identify 16 possible ${\bf Z}_2$ monodromy
patterns (called {\it sectors}) in the case of a closed $N{=}2$ string.
In sect.~4 we use bosonization techniques to introduce spin and twist fields.
These fields are needed to specify the vertex operators creating the ground
states in the 16~twisted sectors.
Next, we calculate the critical intercepts for all these sectors, and exhibit
the possible {\it massless} ground states.
Sect.~5 is devoted to a discussion of the spectral flow in $N{=}2$
string theory. We generalize the boundary conditions to twists by
arbitrary phases and give, again, the relevant formulae for the ground state
energy. The most general monodromy conditions are
the subject of sect.~6, where we classify all possibilities in terms of the
conjugacy classes of $U(1,1)$, and propose the most general intercept formula.
We find a somewhat peculiar new massless background, related to the parabolic
conjugacy class of~$SU(1,1)$.
Sect.~7 deals with the BRST cohomology and interactions in $N{=}2$ string
theory. First, we check the BRST-invariance of the massless candidate
(${\bf Z}_2$-twisted) ground states, and, second, investigate their possible
interactions from the locality requirement for the vertex operator algebra.
Our conclusions are summarized in sect.~8.
Two Appendices comprise auxiliary information about the local transformation
laws of the $N{=}2$ string fields (Appendix~A) and the $N{=}2$ string BRST
charge (Appendix~B).
\vglue.2in

\section{String boundary conditions}

When varying a string action, some boundary terms appear. The action principle
implies that certain constraints are to be added to the string equations of
motion, in order to eliminate the boundary terms.

Varying the gauge-fixed bosonic string action~\cite{gsw}~\footnote{We suppress
target space indices. The dots stand for their contractions with a flat
metric $\h_{\m\n}$.}
$$S_0\ =\ \fracmm{1}{2\p}\int^T_0d\t\int^{\p}_0 d\s\,\h^{\a\b}\pa_{\a}X\cdot
\pa_{\b}X\eqno(2.1)$$
yields not only the equations of motion $(\Box X=0)$, but also the constraints
\footnote{The prime and dot over a function mean the differentiation with
respect to $\s$ and $\t$, respectively.}
$$\eqalign{
{\rm open}: ~~&~~ \left.X'\right|_{\s=0,\p}=0~,\quad
\left.\dt{X}\right|_{\t=0,T}=0~;\cr
{\rm closed}: ~~&~~  X' - ~~{\rm periodic}~,\quad
\left.\dt{X}\right|_{\t=0,T}=0~.}\eqno(2.2)$$
According to the minimal action principle, $\dt{X}$ is supposed to vanish
for the initial and final string states, while the `stringy' condition on
$X'$ is relevant.
More general boundary conditions than the above are possible, but correspond to
identifications in the target space and, hence, change global topology.

In the case of the $N{=}1$ fermionic gauge-fixed string action \cite{gsw}
$$\eqalign{
S_1\ =&\ \fracmm{1}{2\p}\int d\t d\s\,\left\{ \pa_{\a}X\cdot \pa^{\a}X
+\fracmm{i}{2}\Bar{\J}\cdot\r^{\a}\pa_{\a}\J\right\}\cr
=&\ \fracmm{1}{\p}\int d^2\x\,\left\{ \pa_{+}X\cdot \pa_{-}X
+\J_-\cdot\pa_+\J_- + \J_+\cdot\pa_{-}\J_+\right\}~,\cr}\eqno(2.3)$$
where light-cone coordinates $\x_{\pm}=\t\pm\s$ as well as real
Majorana-Weyl fermion fields $\J_{\pm}$ have been introduced on the
world-sheet, one gets, in addition to eq.~(2.2), the fermionic constraints
$$\eqalign{
{\rm open}: ~~&~~ \left.\J_+=\pm\J_-\right|_{\s=0,\p}=0~;\cr
{\rm closed}: ~~&~~ \J_{\pm} - ~~{\rm periodic~or~antiperiodic}~.}\eqno(2.4)$$
The two different possibilities in choosing the signs in eq.~(2.4) are
relevant, and they lead to the known distinction between the Ramond (R) and
Neveu-Schwarz (NS) sectors of the $N{=}1$ string theory \cite{gsw}.
In principle, we are free to choose different boundary conditions for
different target space components of $\J_{\pm}$. This would, however, destroy
global target space symmetries, such as spacetime Lorentz invariance.

In the case of the $N{=}2$ string, there are many more choices for the boundary
conditions since both bosonic and fermionic string world-sheet fields
($N{=}2$ superconformal `matter') become {\it complex}.
The gauge-fixed $N{=}2$
string action has the form $(Z^{\pm}=X\pm iY~,~~\J^{\pm}=\J^1\pm i\J^2)$
$$\eqalign{
S_2\ =&\ \fracmm{1}{\p}\int d^2\x\,\left\{ \pa_{+}Z^{+}\cdot \pa_{-}Z^-
+ \pa_{-}Z^{+}\cdot \pa_{+}Z^-\right.\cr
&\left. +\J^+_+\cdot\pa_-\J^-_+ + \J^+_-\cdot\pa_{+}\J^-_-
+\J^-_+\cdot\pa_-\J^+_+ + \J^-_-\cdot\pa_{+}\J^+_-\right\}~.\cr}\eqno(2.5)$$

Zero modes of the $Z$ fields are normally identified with the coordinates
of embedding `spacetime' in which the $N{=}2$ string propagates.
Both the dimension and
the signature of this `spacetime' are fixed quantum-mechanically, when we
require conformal anomaly cancellation (or nilpotency of the BRST
charge) in the absence of additional propagating degrees of freedom
(like Liouville modes).
Critical $N{=}2$ strings are known to live in two {\it complex}
dimensions \cite{ft} (see refs.~\cite{mar,ket} for the recent reviews about
the $N{=}2$ strings). The $(2,2)$ `spacetime' signature of $N{=}2$ critical
string theory is apparently non-physical. Consequently, there seems to be no
compelling reason
to insist on a direct physical interpretation of the $N{=}2$ string target
space at all, or assume it to be smooth everywhere.
Let us therefore allow as much freedom as possible at this point,
and do not even assume that our flat `spacetime' is globally a manifold.

Of course, we do not have total freedom in the choice of boundary conditions.
They have to be compatible with the symmetries of the action.
More precisely, we demand the action density to be single-valued on the
world-sheet. Being moved around along a closed path, it should come back to
its
original value. The corresponding monodromies of the world-sheet fields must
therefore conspire to produce a symmetry of the action (density).
In order to take into account the constraints which accompany the gauge-fixed
action~(2.5), we shall investigate the monodromies for the full
gauge-invariant action. The monodromies have to respect the local
world-sheet symmetries as well as the global target space symmetries.
Classifying allowed boundary conditions (and thus possible flat backgrounds)
requires an analysis of all (global and local) symmetries.
Apart from this, we do not constrain our world-sheet fields at all.

Stated differently, at the outset we allow a {\it multi-valuedness}
of the $N{=}2$ string coordinates $Z$ and $\J$.
This implies that generic monodromy factors
may appear in the boundary conditions, and each matter field may live
in a quite general twisted complex bundle. Given a Riemann surface~$\S$ to
represent the $N{=}2$ string world-sheet after its `euclideanization', it
is the allowed choice of the {\it global monodromies}, i.e. `phases' picked up
by moving the matter fields around the cycles comprising a homology basis on
$\S$, that will be in question of our discussion in the next sections.

A monodromy matrix~$U$ generates a subgroup~$\G$ of the symmetry group~$G$ of
the string action.
The twisted string target space is therefore obtained from the untwisted
one through modding out by~$\G$.
On the bosonic coordinates~$Z$ some subgroup~$G_0\subset G$ is realized,
and the relevant monodromy creates some~$\G_0\subset G_0$.
Clearly, the {\it bosonic} background~$\cb$ is simply
the quotient space~${\bf C}^{1,1}/\G_0$.
If one likes to retain a (2+2)-dimensional manifold for~$\cb$ locally,
$\G_0$ is required to be discrete but not necessarily finite.
Typical examples are orbifolds from $\G_0={\bf Z}_n$, $n\geq2$,
or tori from $\G_0={\bf Z}$.
The simplest cases are ${\bf Z}_2$ orbifolds,
i.e.~$\cb={\bf C}^{1,1}/{\bf Z}_2$.
For this reason we shall discuss them first, in the two following sections.
We will return to the general situation in sects.~5 and~6.
This means that fields of {\it integral} spin are allowed to pick
up signs, i.e. be {\it double}-valued on $\S$, just as for fields of {\it
half-integral} spin.
The complete monodromy behavior will be fixed from the signs picked up
by the components of $Z$, in addition to an overall sign between $Z$ and $\J$
related to the NS-R distinction.

A natural example of the twisted boundary conditions is
the {\it Mathur-Mukhi} choice considered in ref.~\cite{mm},
$$Z^{\pm}(\p)= Z^{\mp}(0)~,\qquad \J^{\pm}(\p)=-\J^{\mp}(0)~\;
{\rm or}\; {+}\J^{\mp}(0),\eqno(2.6)$$
uniformly with respect to all the (suppressed) target space indices.
This allows us to choose different signs
for real and imaginary parts of the fields, while keeping $X={\rm
Re}\,Z^{\pm}$ to be periodic.
The choice of boundary conditions in eq.~(2.6) is obviously consistent with
the variation of the action (2.5).

The most conservative ({\it Ooguri-Vafa}) choice \cite{ov} of the $N{=}2$
string boundary conditions in the form
$$Z^{\pm}(\p)= Z^{\pm}(0)~,\quad
\J^{\pm}(\p)=({+}\,{\rm or}\,{-})\J^{\pm}(0)~, \eqno(2.7)$$
is the only one which allows us to keep the single-valuedness of integral spin
fields, and ${\bf C}^{1,1}$ as the consistent $(2{+}2)$-dimensional background
`spacetime' for $N{=}2$ string propagation. This choice only deals with
untwisted line bundles and their square roots (spin bundles) to define
fermions, just like for the $N{=}1$ string.
The two possible
signs in eq.~(2.7) are common for all the world-sheet spinors, and
correspond to the usual NS-R distinction familiar from the $N{=}1$ case.

The two boundary conditions presented so far are blind to the `spacetime'
indices of $Z$ and $\J$ and thus compatible with naive real `spacetime'
Lorentz symmetry~$O(1,1)$. This is also true for an
additional overall sign flip in eqs. (2.6) and~(2.7), which doubles the
number of such cases to four.~\footnote{
For a genus-$h$ Riemann surface $\S$, there
are $4^{2h}$ possible boundary conditions of this type
for the $Z$ fields alone.} It is
conceivable that we might finally need to sum over all backgrounds or,
equivalently, over all `spin structures' in the $N{=}2$ string partition
function. The only compelling reason to do so might possibly come from
modular invariance, since the twists imply drastic consequences for the $N{=}2$
moduli space. To address this issue, one needs a better understanding
of $N{=}2$ moduli space, which is a rather involved problem.

Before going any further, we want to briefly discuss our notation.
Target space indices (internal and Lorentz) always appear as superscripts,
world-sheet indices usually as subscripts.
In real components
$$Z^{\m\pm}=Z^{\m 2}\pm i Z^{\m 3}\quad,
\quad\J^{\m\pm}=\J^{\m 2}\pm i\J^{\m 3}~,\eqno(2.8)$$
we have the fields $Z^{\m i}(\x)$ and
$\J^{\m i}(\x)$,  with generic monodromy conditions of the form
$$Z^{\m i}(\p)=M\ud{\m i}{\n j}Z^{\n j}(0)~,\eqno(2.9)$$
and similarly for the $\J$'s. The lower-case Greek indices $\m,\n=0,1$ refer
to a 2-dimensional Minkowski space of signature $(-,+)$, while the
lower-case Latin indices $i,j=2,3$ refer to the real components of the complex
fields. To avoid confusing the same numerical values of $\m$ and $i$,
we have taken a slightly unusual range for the lower-case Latin indices,
so that we have
$$\ula \equiv \{\m i\}=(02,03,12,13)~.\eqno(2.10)$$

Sometimes, we write shorthand $Z^\m$ for $(Z^{\m 2},Z^{\m 3})$ or
$(Z^{\m +},Z^{\m -})$, as well as $Z^i$ for $(Z^{0i},Z^{1i})$ or
$Z^\pm$ for $(Z^{0\pm},Z^{1\pm})$, and similarly for $\J$, suppressing
irrelevant indices in an obvious fashion.
It may also be convenient to choose spacetime light-cone coordinates
$Z^{\ua\da}=Z^0{\ua\da}Z^1$, where $\ua\da$ is just another set of $\pm$ signs
(index~$i$ is suppressed here.)

In the $N{=}2$ superconformal gauge, the fields (and ghosts) of the $N{=}2$
string on the euclidean world-sheet become free, so that they can be
decomposed into their holomorphic and anti-holomorphic parts, as is usual
in two-dimensional conformal field theory. For definiteness, we investigate
only {\it closed} $N{=}2$ strings in this paper.
If not mentioned explicitly, we will consider only the right-movers in the
following. As usual, lower-case letters will be used to denote the chiral
parts of the matter fields, except for the $Z$ fields
(see Appendix B for more details).
\vglue.2in

\section{Twisting the $N{=}2$ string}

The world-sheet action of $N{=}2$ fermionic string theory reads \cite{bs}
\footnote{We use a purely imaginary (Majorana) representation for
two-dimensional Dirac
matrices $\r^{\a}$, $\a=0,1$: $\r^0=\s_2$, $\r^1=i\s_1$, $\r^3=\r^0\r^1=\s_3$,
where $\s_i$, $i=1,2,3$, are Pauli matrices,
and $\{\r^\a,\r^\b\}=-2\h^{\a\b}$, with $\h=diag(-,+)$.
The Majorana spinors $\J=(\J_-,\J_+)$ decompose into their Weyl parts $\J_\pm$.
Dirac conjugation (denoted by
bar) $\Bar{\J}\equiv\J^{\rm T}\r^0$ does {\it not} include complex
conjugation (denoted by superscript `$-$').
Complex conjugation is explicit in our formulae, and it always acts first.}
$$\eqalign{ S\ =&\
\fracmm{1}{\p}\int d^2\x\,e\left\{\ha h^{\a\b}\pa_{\a}Z^+\cdot\pa_{\b}Z^-
+\fracmm{i}{2}\Bar{\J}^+\cdot\r^{\a}\dvec{D}_{\a}\J ^-
+ A_{\a}\Bar{\J}^+\cdot\r^{\a}\J^-\right.\cr
&\left. + \left(\pa_{\a}Z^++\Bar{\J}^+\c^-_{\a} \right)
\cdot\Bar{\c}^+_{\b}\r^{\a}\r^{\b}\J^- + \left(\pa_{\a}Z^-+\Bar{\c}^+_{\a}\J^-
\right)\cdot\Bar{\J}^+\r^{\b}\r^{\a}\c^-_{\b}\right\}~,\cr}\eqno(3.1)$$
where the $N{=}2$ world-sheet supergravity multiplet, comprising a real
zweibein~$e^a_{\a}(\x)$ with the metric $h_{\a\b}=\h_{ab}e^a_{\a}e^b_{\b}$,
a complex gravitino field~$\c^\pm_{\a}(\x)$, and a real $U(1)$ gauge
field~$A_{\a}(\x)$, has been introduced.

The  Brink-Schwarz action (3.1) is known to be invariant under the following
{\it local} world-sheet symmetries:
\begin{itemize}
\item reparametrization invariance,
\item Lorentz invariance,
\item $N{=}2$ supersymmetry,
\item phase and chiral $U(1)$ gauge invariances,
\item Weyl and super-Weyl invariances.
\end{itemize}
The explicit form of all the local transformation laws in components can be
found in Appendix A (see ref.~\cite{km} for the $N{=}2$ superspace
description of the $N{=}2$ string action and its symmetries). In particular,
the local $U(1)$ symmetry acts non-trivially on the fields
$\J^\pm$, $\c^\pm_{\a}$ and $A_{\a}$ only.

The action (3.1) also possesses {\it global continuous} invariances
associated with target space symmetries, namely
\begin{itemize}
\item global translation invariance ${\bf C}^{1,1}$,
\item global $U(1)$ symmetry acting on the internal indices of the matter
fields $Z^i$ and $\J^i$, with complex eigenstates
$Z^\pm=Z^2 \pm iZ^3$ and $\J^\pm=\J^2 \pm i\J^3$,
\item global $O(1,1)$ `Lorentz' symmetry acting on the `spacetime'
indices of the matter fields $Z^{\m}$ and $\J^{\m}$, with light-cone
eigenstates
$Z^{\ua\da}=Z^0{\ua\da}Z^1$ and $\J^{\ua\da}=\J^0{\ua\da}\J^1$.
\end{itemize}
The intersection of the two symmetries is a ${\bf Z}_2$ symmetry
generated by the total sign flip~PT of all matter fields, where P and T
denote the usual `parity' and `time reversal' transformations, respectively.

The full global continuous symmetry of the action is, however, still larger,
since the $O(1,1)$ symmetry can be extended to
$U(1,1) = \left[ U(1)\otimes SU(1,1)\right]/{\bf Z}_2$,
where the factors $U(1)$ and ${\bf Z}_2$ coincide with the ones
just mentioned. More precisely,
we may split $O(1,1)=SO(1,1)\otimes{\bf Z'}_2$, with ${\bf Z'}_2$ generated by
the `parity'~P, and add complex boost and rotation generators to create
$SU(1,1)$. The ${\bf Z'}_2$ is then contained in the product with the $U(1)$
symmetry already present.

In 2+2 dimensions, the natural `spacetime' Lorentz symmetry for the
$N{=}2$ string is
$SO(2,2)=\left[ SU(1,1)\otimes SU(1,1)\right]/{\bf Z}_2$ ~\cite{ket}.
The interaction terms in the action (3.1) break, however,
one of the two $SU(1,1)$
factors down to $U(1)\otimes{\bf Z''}_2$, with ${\bf Z''}_2$ representing
the Mathur-Mukhi twist (sect.~2).
It has been conjectured \cite{siegel} that the total global Lorentz symmetry of
the $N{=}2$ string in the given formulation might actually be the remnant of a
`hidden' $SO(2,2)$ symmetry, which presumably exhibits itself in the
equivalent $N{=}4$ supersymmetric formulation of the same string theory.

Finally, the $N{=}2$ string action (3.1) is invariant under the four different
${\bf Z}_2$ twists of the fields:
$$\eqalign{
{\bf Z}_2~: ~~&~~ Z\to -Z\qquad{\rm and~the~same~for}\quad\J~,\cr
{\bf Z'}_2~: ~~&~~ Z^0\to Z^0~,~~Z^1\to -Z^1
\qquad{\rm and~the~same~for}\quad\J^{0,1}~,\cr
{\bf Z''}_2~: ~~&~~ Z^\pm\to Z^\mp\qquad{\rm and~the~same~for}\quad
\J^\pm~,\cr
{\bf Z'''}_2~: ~~&~~ Z\to Z\qquad{\rm and}\qquad\J\to -\J~,\cr}\eqno(3.2)$$
by adjusting the behavior of the $N{=}2$ supergravity fields appropriately
(see below).
These {\it global discrete} symmetries are imbedded in the continuous groups
discussed above. In particular,
${\bf Z}_2=O(1,1)\cap U(1)=SU(1,1)\cap U(1)$ inverts (PT), whereas
${\bf Z'}_2$ is a `large' transformation (P) in $O(1,1)$.
The twist ${\bf Z''}_2\subset SO(2,2)$ is the only one which is
not contained in $U(1,1)$.
The ${\bf Z'''}_2$ twist is part of the local $U(1)$ and does not affect $Z$;
it apparently resembles the usual NS$\leftrightarrow$R
twist in the $N{=}1$ string theory.
In the special case of ${\bf Z}_2$ monodromies we are therefore
going to use the same labels
(NS and R) to distinguish these two cases, even though their actual meaning
is quite different in the $N{=}2$ string theory.
Generally speaking, the NS (R) sector is characterized
by $\J^{\m i}$ and $Z^{\m i}$
having opposite (identical) monodromy signs.
In summary, the total homogeneous {\it global} symmetry group
of the $N{=}2$ string is $G_0=U(1,1)\otimes{\bf Z''}_2$,
which is at the same time the maximal global
monodromy group for the matter fields $Z$ or $\J$, seperately.
The {\it relative} boundary conditions between $Z$ and $\J$ can only arise
from {\it local} symmetries, so that $Z$ and $\J$ monodromies are
rigidly related after gauge fixing.

It is not difficult to deduce the consequences of eq.~(3.2) for the $N{=}2$
supergravity fields. First, one easily sees from inspecting the gravitino
couplings in the action (3.1) that the complex gravitino field
$\c^\pm_{\a}$ should
transform under each twisting in exactly the {\it same} way as the complex
matter field $\J^\pm$, uniformly for each $\a$ value.
Under ${\bf Z''}_2$, for example,
$$\eqalign{ \left(\pa_{\a}Z^++\Bar{\J}^+\c^-_{\a}\right)\
&\stackrel{\rm twist}{\longrightarrow}\
\left(\pa_{\a}Z^- + \Bar{\J}^-\tilde{\c}^-_{\a}\right)\
=\ \pa_{\a}Z^- + \Bar{\tilde{\c}}^-_{\a}\J^-~,\cr
\Bar{\c}^+_{\b}\r^{\a}\r^{\b}\J^-\
&\stackrel{\rm twist}{\longrightarrow}\
\Bar{\tilde{\c}}^+_{\b}\r^{\a}\r^{\b}\J^+\
=\ \Bar{\J}^+\r^{\b}\r^{\a} \tilde{\c}^+_{\b}~,\cr}\eqno(3.3)$$
where the tilde over a field means its ${\bf Z''}_2$ twisting. Similarly,
$$A_{\a}\Bar{\J}^+\cdot\r^{\a}\J^-\ \stackrel{\rm twist}{\longrightarrow}\
\tilde{A}_{\a}\Bar{\J}^-\cdot\r^{\a}\J^+\
=\ -\tilde{A}_{\a}\Bar{\J}^+\cdot\r^{\a}\J^-~.\eqno(3.4)$$
The minus on the r.h.s. of eq.~(3.4) is important, since it implies
$$A_{\a}\ \stackrel{\rm twist}{\longrightarrow}\ -A_{\a}\eqno(3.5)$$
under the ${\bf Z''}_2$ twist.

For the {\it untwisted} boundary conditions one gets
$$\eqalign{
Z(\p)=Z(0)~, ~~&~~ A_{\a}(\p)=A_{\a}(0)~,\cr
\J(\p)=\pm\J(0)~, ~~&~~ \c_{\a}(\p)=\pm\c_{\a}(0)~,\cr}\eqno(3.6)$$
where the signs in the second line are in correspondence with each other
and denote additional optional ${\bf Z'''}_2$ twist, {\it i.e.} the NS/R
option. For the ${\bf Z''}_2$ {\it twisted} boundary conditions one has instead
$$\eqalign{
Z^+(\p)=Z^-(0)~, ~~&~~ A_{\a}(\p)=-A_{\a}(0)~,\cr
\J^+(\p)=\pm\J^-(0)~, ~~&~~ \c^+_{\a}(\p)=\pm{\c^-}_{\a}(0)~.\cr}\eqno(3.7)$$
Eq.~(3.7) implies, in particular, that the (Abelian) first Chern class
$c=\fracmm{1}{2\p}\int_{\S}F$, $F\equiv dA$, associated with the $U(1)$ gauge
field $A$ on a Riemann surface $\S$, has to vanish for such twisted boundary
conditions.

The monodromy group $U(1,1)\otimes{\bf Z''}_2$ does not have room
for all 16 independent sign choices of the four $Z^{\m i}$ (or $\J^{\m i}$)
components. From the possibilities listed in eq.~(3.2) it is clear
that in each ${\bf Z}_2$-twisting
there must always be an {\it even} number of minuses among the four
components~$(\m i)$,~\footnote{Let us call these twistings {\it even}.}
in order for the compensating monodromy of the supergravity fields
$\c^\pm_{\a}$ and $A_{\a}$ to exist.
For instance, flipping only {\it one} component of $\J$, say
$\J^{02}$, is inconsistent with the $N{=}2$ string action,
as can easily be seen from the term
$$A_{\a}\;\Bar{\J}^+{\cdot}\r^{\a}\J^-\ =\ 2iA_{\a}\left(
\Bar{\J}^{02}\r^{\a}\J^{03}-\Bar{\J}^{12}\r^{\a}\J^{13}\right)~.
\eqno(3.8)$$

We have checked the compatibility of any even twisting with all the local
symmetries of the $N{=}2$ string action (3.1), listed in Appendix~A.
The procedure is straightforward,
and it simply determines the behavior of the local symmetry parameters under
the twisting.
The (superconformal) gauge fixing in $N{=}2$ string theory results in the
$N{=}2$ superconformal algebra in terms of the currents associated
with the residual symmetries. The related ghost structure and the BRST charge
are well known \cite{mm,bilal,bien}, and the  results are collected in
Appendix B. Naturally, the boundary conditions of the currents and, hence,
their moding, depend on the twisting, e.g. eq.(3.7) implies an
antiperiodic $U(1)$ current~$J$.

An abelian subgroup of the full symmetry group is
$U(1)\otimes SO(1,1)\otimes U(1)_{\rm gauged}$, where the first two factors
represent global symmetries. Let $(q,s,e)$ be the corresponding charges of
various fields (including ghosts) with respect to $U(1)$, $SO(1,1)$ and
$U(1)_{\rm gauged}$, respectively.
Their complete list is compiled in Table~I.
\vglue.2in

\noindent{\sf Table I}.
The charges  $(q,s,e)$ and conformal dimensions~$h$ of the world-sheet fields
with respect to $U(1)$, $SO(1,1)$, $U(1)_{\rm gauged}$ and $Vir$, respectively.
Blanks appear where the fields are not eigenstates.
\vglue.1in
\noindent\begin{tabular}{c|c|c|c|c|c|c|c|c|c|c|c|c|c} \hline
field & $Z^\pm$ & $Z^{\ua\da}$ & $Z^{\ua\da\pm}$ &
$\J^\pm$ & $\J^{\ua\da}$ & $\J^{\ua\da\pm}$ &
$t^\pm$ & $t^{\ua\da}$ & $S^\pm$ & $S^{\ua\da}$ &
$\c_\a^\pm$ & $\b^\pm$ & $\g^\pm$ \\ \hline
$q$ & $\pm1$ & & $\pm1$ & $\pm1$ & & $\pm1$ & $\pm\ha$ & & $\pm\ha$ & &
0 & 0 & 0 \\
$s$ &  & ${\ua\da}1$ & ${\ua\da}1$ & & ${\ua\da}1$ & ${\ua\da}1$ & &
${\ua\da}\fracmm{1}{2}$ & & ${\ua\da}\fracmm{1}{2}$ &
0 & 0 & 0 \\
$e$ & 0 & 0 & 0 & $\pm1$ & & $\pm1$ & 0 & 0 & $\pm\fracmm{1}{2}$ & &
$\pm1$ & $\pm1$ & $\pm1$ \\ \hline
$h$ & 0 & 0 & 0 & $\fracmm{1}{2}$ & $\fracmm{1}{2}$ & $\fracmm{1}{2}$ &
$\fracmm{1}{8}$ & $\fracmm{1}{8}$ & $\fracmm{1}{8}$ & $\fracmm{1}{8}$ &
& $\fracmm{3}{2}$ & $-\fracmm{1}{2}$ \\ \hline
\end{tabular}
\vglue.2in
\noindent
The rest of the
$N{=}2$ supergravity fields, $(e^a_{\a},A_{\a})$, the reparametrization ghosts
$(b,c)$ and the $U(1)$ ghosts $(\tilde{b},\tilde{c})$ are all inert with
respect to $U(1)\otimes SO(1,1)\otimes U(1)_{\rm gauged}$.

The monodromy conditions for all the  ${\bf Z}_2$-twisted sectors of the
closed $N{=}2$ string are collected in Table II, where pluses and minuses
mean periodicity and anti-periodicity, respectively.
There are $2^4{=}16$ sectors in total, due to the ${\bf Z}_2^4$ twist group.

\newpage

\noindent {\sf Table II}.
The ${\bf Z}_2^4$ monodromy conditions for the (twisted) sectors of the $N{=}2$
closed string (right movers only).
The last rows give conformal dimensions and local $U(1)$ charges
of the associated zero-momentum ground states (sect.~4).
\vglue.1in
\noindent\begin{tabular}{c|cc|cc|cc|cc|cc|cc|cc|cc} \hline
\# & 1 & 2 & 3 & 4 & 5 & 6 & 7 & 8 & 9 & 10 & 11 & 12 & 13 & 14 & 15 & 16 \\
\hline
${\bf Z'''}_2$ & R & NS & R & NS & R & NS & R & NS & R & NS & R & NS & R & NS
& R & NS \\ \hline
$Z^{02}$ & + & + & + & + & $-$ &  $-$ & $-$ & $-$ & + & + & + & +
& $-$ & $-$ &  $-$ &  $-$ \\
$Z^{03}$ & + & + & + & + &  $-$ &  $-$ &  $-$ &  $-$ & $-$ & $-$ & $-$ & $-$
& + & + & + & + \\
$Z^{12}$ &  + & + &  $-$ &  $-$ & + & + & $-$ & $-$ & + & + &  $-$ &  $-$
& + & + & $-$ & $-$ \\
$Z^{13}$ & + & + &  $-$ &  $-$ & + & + &  $-$ & $-$  & $-$ & $-$ & + & +
& $-$ & $-$ & + & + \\
\hline
$\J^{02}$ &  + &  $-$ & + &  $-$ &  $-$ & + & $-$ & + & + & $-$ & + & $-$
& $-$ & + &  $-$ & + \\
$\J^{03}$ & + &  $-$ & + &  $-$ &  $-$ & + &  $-$ & + & $-$ & + & $-$ & +
& + & $-$ & + &  $-$ \\
$\J^{12}$ & + &  $-$ &  $-$ & + & + &  $-$ & $-$ & + & + & $-$ &  $-$ & +
& + &  $-$ & $-$ & + \\
$\J^{13}$ & + &  $-$ &  $-$ & + & + &  $-$ &  $-$ & + & $-$ & + & + & $-$
& $-$ & + & + & $-$ \\
\hline
$\c^2$ & + &  $-$ & + &  $-$ & + &  $-$ & + &  $-$ & + & $-$ & + & $-$ & + &
 $-$ & + & $-$ \\
$\c^3$ & + &  $-$ & + &  $-$ & + &  $-$ & + & $-$ & $-$ & + &  $-$ & + &  $-$
 & + & $-$ & + \\
\hline
$A_{\a}$ & + & + & + & + & + & + & + & + &  $-$ &  $-$ & $-$ & $-$ & $-$ &
 $-$ &  $-$ & $-$  \\
\hline
$h$ & 0 & 0 & 0 & $\fracmm{1}{4}$ & 0 & $\fracmm{1}{4}$ & 0 & $\fracmm{1}{2}$
& 0 & 0 & 0 & 0 & 0 & 0 & 0 & 0 \\
\hline
$e$ & 0 & 0 & $\pm\fracmm{1}{2}$ & $\pm\fracmm{1}{2}$ & $\pm\fracmm{1}{2}$
& $\pm\fracmm{1}{2}$ & 0 & 0 & - & - & - & - & - & - & - & - \\ \hline
\end{tabular}
\vglue.2in
\vglue.2in

\section{Bosonization, spin fields and twist fields}

The gauge-fixed action~(2.5) is accompanied by a ghost action.
It is well-known~\cite{mar,ket} that the ghost systems appropriate for the
$N{=}2$ string are
\begin{itemize}
\item the reparametrization ghosts ($b,c$), an anticommuting pair of
free world-sheet fermions with conformal dimensions~($2,-1$).
\item the $N{=}2$ supersymmetry ghosts ($\b^i,\g^i$) or ($\b^\pm,\g^\pm$),
two commuting pairs of free world-sheet fermions with conformal
dimensions~($\frac32,-\frac12$).
\item the $U(1)$ ghosts ($\tilde{b},\tilde{c}$), an anticommuting pair of
free world-sheet fermions with conformal dimensions~($1,0$).
\end{itemize}

Since the superconformal ghosts $(\b^i,\g^i)$ carry a `spacetime' index $i$
associated with the target space where the $N{=}2$ string lives, the
`spacetime' properties of the physical $N{=}2$ string states (to be determined
by the BRST cohomology) may depend on their ghost structure as well.
This would clearly be
quite different from the conventional cases of the bosonic and $N{=}1$
supersymmetric strings. To settle the framework for identifying the
${\bf Z}_2$-twisted BRST-invariant states,
we have to build up the ground states for all sectors (given
in Table II) out of the single $N{=}2$ super-$SL(2,{\bf C})$ invariant
vacuum $\ket{0}$. This can be done by introducing spin and
${\bf Z}_2$ twist fields for the matter and ghost fields,
and using bosonization \cite{fms,k-w,gom}. The spin
fields twist world-sheet fermions, while the twist fields twist world-sheet
bosons. From now on we employ planar complex coordinates $(z,\bar z)$ for the
euclidean world-sheet.

The chiral fermionic fields $\j^{\m i}(z)$ can be bosonized as~\footnote{Normal
ordering is always suppressed, as well as cocycle operators~\cite{k-w}.}
$$\j^{0i}\cong \fracmm{1}{\sqrt{2}}\left(e^{\ua\f^i}+e^{\da\f^i}\right)~,\qquad
\j^{1i}\cong\fracmm{1}{\sqrt{2}}\left(e^{\ua\f^i}-e^{\da\f^i}\right)~,
\eqno(4.1)$$
where $\ua$ and $\da$ simply stand for ${+}$ and ${-}$.
In eq.~(4.1) two real scalar bosons $\f^i(z)$ have been introduced,  with the
operator product expansion (OPE)
$$\f^i(z)\ \f^j(w)\ \sim\ \d^{ij}\ln (z-w)~.\eqno(4.2)$$
Using the light-cone combinations
$$\fracmm{1}{\sqrt{2}}\,\j^{\ua\da i}\cong e^{\ua\da\f^i}~,\eqno(4.3)$$
we can construct the {\it spin} fields $S^{i,\ua\da}(z)$
with helicity index $\ua\da$ as
$$S^{i,\ua\da}\cong e^{\ua\da\f^i/2}~.\eqno(4.4)$$
We choose the notation $\ua\da$ for the `spacetime' light-cone coordinates in
order to distinguish them from the complex internal ones `$2\pm i3$'.
The spin fields $S^{i,\ua\da}$ twist the fermions $\j^{\m i}$ with respect to
the index $\m$, which corresponds to the action of ${\bf Z''}_2$ and the
$(+-+-)$-type boundary conditions in Table~II.

The convenient application of the bosonization procedure actually depends
on the monodromy conditions under consideration. If the twist
is applied with respect to the index $i$, which corresponds to the
action of  ${\bf Z'}_2$ and the
$(++--)$-type boundary conditions in Table~II, different
bosonization rules have to be introduced as \cite{gom}
$$\j^{\m 2}\cong \fracmm{i^\m}{\sqrt{2}}\left(e^{\f^{\m}}+e^{-\f^{\m}}\right)~,
\qquad \j^{\m 3}\cong \fracmm{i^\m}{i\sqrt{2}}\left(e^{\f^{\m}}-e^{-\f^{\m}}
\right)~,\eqno(4.5)$$
with the OPE as
$$\f^\m(z)\ \f^\n(w)\ \sim\ \d^{\m\n}\ln (z-w)~.\eqno(4.6)$$
The associated $U(1)$ and $U(1)_{\rm gauged}$ eigenfields are
$$\fracmm{1}{\sqrt{2}}\,\j^{\m\pm}\cong i^\m\,e^{\pm\f^\m}\quad,\quad
S^{\m,\pm}\cong e^{\pm\f^\m/2}~.\eqno(4.7)$$
The $(q,s,e)$ charges of any spin field can be found in Table~I.
We can also formally define a `little' spin field~$S^{\m i}$ which simply
sign-flips the boundary condition of the individual fermion~$\j^{\m i}$
and has conformal dimension $h=\fracmm{1}{16}$.

The reparametrization ghosts $(b,c)$ are bosonized as \cite{k-w}
$$b\cong e^{-\s}~,\quad c\cong e^{\s}~,
\qquad{\rm with}\quad \s(z)\ \s(w)\ \sim\ \ln (z-w)~.\eqno(4.8)$$
Similarly, for the $U(1)$ ghosts $(\tilde{b},\tilde{c})$ one has
$$\tilde{b}\cong e^{-\tilde{\s}}~,\quad \tilde{c}\cong e^{\tilde{\s}}~,
\qquad{\rm with}\quad \tilde{\s}(z)\ \tilde{\s}(w)\ \sim\ \ln (z-w)~.
\eqno(4.9)$$
As for the superconformal ghosts $(\b^i,\g^i)$, the bosonization rules
of refs.~\cite{fms,k-w} imply
$$\b^i(z)\cong e^{-\vf^i}\pa \x^i\cong e^{-\vf^i+\q^i}\pa\q^i~,\qquad
\g^i(z)\cong \h^i e^{\vf^i} \cong e^{-\q^i + \vf^i}~,$$
$$ \vf^i(z)\ \vf^j(w)\ \sim\ -\d^{ij}\ln (z-w)~,
\qquad \h^i(z)\ \x^j(w)\ \sim\ \d^{ij}\fracmm{1}{z-w}~,\eqno(4.10)$$
where the auxiliary $(\h^i,\x^i)$ conformal system of spin (1,0)
has also been bosonized as
$$\h^i\cong e^{-\q^i}~,\quad \x^i\cong e^{\q^i}~,\qquad{\rm with}\quad
\q^i(z)\ \q^j(w)\ \sim\ \d^{ij}\ln (z-w)~.\eqno(4.11)$$
The `solitons' $e^{\pm\vf^i}$ are outside the monomial field algebra of
$(\b^i,\g^i)$; however, one finds that
$$e^{\vf^i}\cong \d(\b^i)~,\qquad e^{-\vf^i}\cong \d(\g^i)~.\eqno(4.12)$$
Alternatively, one can bosonize the complex linear combinations
$\b^\pm=\b^2\pm i\b^3$ and $\g^\pm=\g^2\pm i\g^3$,
$$\b^{\pm}\cong e^{-\vf^{\mp}}\pa\x^{\pm}~,\quad
\g^{\pm}\cong \h^{\pm} e^{\vf^{\pm}}~,\quad
\d(\b^{\pm})\cong e^{\vf^\mp}~,\quad \d(\g^{\pm})\cong e^{-\vf^{\pm}}~,$$
$$\vf^{\pm}(z)\ \vf^{\pm}(w)\ \sim\ -\ln(z-w)~,
\qquad \h^\pm(z)\ \x^\mp(w)\ \sim\ \fracmm{2}{z-w}~,\eqno(4.13)$$
where $\h^\pm$ and $\x^\pm$ are not just linear combinations of the fields
appearing in eq.~(4.11).

Here and in what follows we use the standard results \cite{fms,k-w}
for free chiral bosons $\r\in\{\s,\tilde\s,\vf^i,\q^i\}$
$$h\left[e^{q\r}\right]\ =\ \fracmm{\e}{2}q(q-\tilde{Q})~,\qquad
\r(z)\ \r(w)\ \sim\ \e\ln(z-w)~,\qquad \e=\pm 1~,$$
$$\tilde{Q}_\s=3~,\qquad \tilde{Q}_{\tilde\s}=1~,\qquad
\tilde{Q}_{\vf^i}=-2~,\qquad \tilde{Q}_{\q^i}=1~,\eqno(4.14)$$
where $\tilde{Q}$ is a background charge, and the factor $\e$ takes into
account statistics. It follows that
the spin fields $S^{i,\ua\da}$ twisting $\j^i$ and $\c^i$, as well as the
spin fields $S^{\m,\pm}$ twisting $\j^{\m}$, have conformal dimensions
equal to $+\fracmm{1}{8}$.  Similarly, the fields $e^{-\vf^i/2}$
twisting $\b^i$ and $\g^i$ as well as the fields $e^{-\vf^\pm/2}$
twisting $\b^\pm$ and $\g^\pm$ have conformal dimension equal
to~$+\fracmm{3}{8}$.

To describe the possible $N{=}2$ string ground states corresponding to all the
sectors listed in Table~II, we start from the formal vacuum state
$\ket{0}$ in the untwisted sector~(2), which satisfies the
standard constraints ($n\in{\bf Z}$, $r\in{\bf Z}+\ha$)~\footnote{
The mode expansions of the fields and currents
for this case are given in Appendix B.}
$$\eqalign{
\a^{\ula}_n\ket{0}=0\qquad n\geq 0~,&
\qquad\qquad\j^{\ula}_r\ket{0}=0\qquad r\geq 1/2~,\cr
b_n\ket{0}=0 \qquad n\geq -1~,&
\qquad\qquad c_n\ket{0}=0 \qquad n\geq 2~,\cr
\b^i_r\ket{0}=0\qquad r\geq -1/2~,&
\qquad\qquad \g^i_r\ket{0}=0 \qquad r\geq 3/2~,\cr
\tilde{b}_n\ket{0}=0 \qquad n\geq 0~,&
\qquad\qquad\tilde{c}_n\ket{0}=0\qquad n\geq 1~.\cr}
\eqno(4.15)$$
The conformal dimension of the state $\ket{0}$ is $h=0$,
and all its charges vanish, by definition.
Like in the $N{=}1$ string, the BRST cohomology comes in an infinite number
of copies labelled by picture numbers.
More precisely, there are {\it two\/} picture-changing operators
for the $N{=}2$ string, namely
$X^i(z)=\{Q_{\rm BRST},\x^i(z)\}$.
On top of this, there is a four-fold degeneracy represented by
$$e^\s\ket{0}~,\quad e^{2\s}\ket{0}~,\quad e^{\s+\tilde\s}\ket{0}~,
\quad e^{2\s+\tilde\s}\ket{0} \eqno(4.16)$$
in the canonical ghost sectors.~\footnote{
If the $U(1)$ gauge field is twisted, there is no $\tilde c$ zero mode.
In this case, only a two-fold degeneracy arises, and we shall need
$q_{\tilde\s}=\fracm12$.}
We pick the first member of this list here.
Concerning the superconformal ghost pictures, we introduce the notation
$$ \ket{k;q_2,q_3}\ \cong\  e^{\frac{i}2(k^+\cdot Z^- +k^-\cdot Z^+)}
\,e^{q_2\vf^2 + q_3\vf^3 +\s}(0)\ket{0}~\eqno(4.17)$$
and concentrate on the canonical values $q_i\in\{-1,-\ha\}$.
In the following, $\ch^{\rm NS,R}_{\pm\pm\pm\pm}$ denote the 16 sectors of
the complete Fock space of states, with the 4 signs corresponding to the
signs for $Z^{\m i}$ appearing in each column of Table~II.

The reference state $\ket{0}$ is not BRST invariant.
Our representative of the true
$\ch^{\rm NS}_{++++}$ ground state with momentum $k$ is given by
$\ket{k,-1,-1}$, which has $q=s=e=0$ and $h=\ha k^+{\cdot}k^-$,
as can easily be computed from eq.~(4.14).
BRST invariance requires $h=0$ which
translates into the masslessness condition $k^+{\cdot}k^-=0$.
In summary, one finds a massless neutral physical `spacetime' scalar boson.

Among the candidates for the $\ch^{\rm R}_{++++}$ ground state there are
four states of the type
$S^{2,\ua\da}S^{3,\ua\da}\ket{k;-\ha,-\ha}$.
Requiring the ground state to be a $U(1)_{\rm gauged}$-singlet
leaves us with two possibilities,
$$S^{2,\ua}S^{3,\ua}\ket{k;-\ha,-\ha}\qquad{\rm and}\qquad
S^{2,\da}S^{3,\da}\ket{k;-\ha,-\ha}~,\eqno(4.18)$$
which have all charges vanishing and $h=0$, provided
$k^+{\cdot}k^-=0$ again.
In contrast to the NS sector, we identify a neutral massless
physical vector.~\footnote{
The existence of this state in the spectrum of the $N{=}2$
string was noticed in ref.~\cite{gom}.}
One may argue, however, that spectral flow moves us from the
$\ua$~component of this vector to the $\ch^{\rm NS}_{++++}$ ground state
scalar and further to the $\da$~component, effectively identifying
all three degrees of freedom (in the untwisted theory).
In sect.~7 we will test for their BRST invariance.

To actually describe the twisted sectors, we need ${\bf Z}_2$ {\it twist}
fields
$t^{\m i}(z)$, whose role is to twist the boundary conditions for the bosonic
$Z$ fields \cite{za}. The twist fields generically act as
$$t^{\m i}(z)\ \pa Z^{\n j}(w)\ \sim\ \fracmm{\h^{\m\n}\ \d^{ij}}{\sqrt{z-w}}
\,\tilde{t}(w)~,\eqno(4.19)$$
where $\tilde{t}$ is another (regular) twist field. These twist fields act
trivially on $\j$'s.
The conformal dimension of the twist field, $h\left[t^{\m i}\right]=
\fracmm{1}{16}$, and the OPE
$$t^{\m i}(z)\ t^{\n j}(w)\ \sim\ -\fracmm{\h^{\m\n}\ \d^{ij}}{(z-w)^{1/8}}
\eqno(4.20)$$
were calculated in ref.~\cite{dfms}. The results of  ref.~\cite{dfms} imply
conformal dimensions equal to $+\fracmm{1}{8}$ for
the twist fields $t^i\equiv t^{0i}t^{1i}$ (no sum!) and $t^{\m}\equiv t^{\m 2}
t^{\m 3}$ (no sum!), while the dimension of $t\equiv t^{02}t^{03}t^{12}t^{13}$
is equal to~$+\fracmm{1}{4}$.
This is consistent with the change in ground state energy due to a general
$U(1)$ twist, as will be seen in eq.~(5.3) for~$a=\ha$.
We assume that all twist fields are Virasoro primaries.
They carry global $U(1)$ and $SO(1,1)$ charges but are neutral
under $U(1)_{\rm gauged}$ (see Table~I).

Using the spin and twist fields just introduced allows us to construct the
ground states for the
${\bf Z}_2$ twisted sectors of the $N{=}2$ string (Table~II). For example,
for the $\ch^{\rm NS}_{++--}$ sector one finds the doublet
$$t^1 S^{1,\pm} \ket{k;-1,-1}~,\eqno(4.21)$$
where the twist field $t^1$ has been used to twist the $Z^{1i}$,
while the spin field $S^1$ was needed to twist the $\j^{1i}$.
The superconformal ghosts $(\b^i,\g^i)$ are untwisted.
The conformal dimensions of the fields contributing to
this state add up to~$\fracmm14$, and its
local $U(1)$ charge equals~$\pm\fracmm12$.
These properties will be in conflict with BRST invariance.

The candidate ground state for the $\ch^{\rm NS}_{+-+-}$ sector
takes the form \cite{gom}
$$t^3S^{3,\ua\da}e^{\tilde{\s}/2}\ket{k;-1,-\ha}~. \eqno(4.22)$$
Since  the superconformal ghosts $(\b^3,\g^3)$ (and, in fact, $\c^3$)
have to be twisted too, one needs the $q_3=-\ha$ picture
in eq.~(4.22). Finally, the appearance of another `spin' field
$e^{\tilde{\s}/2}$ in eq.~(4.22) is due to the twist of the $U(1)$
gauge field $A_{\a}$ (see Table~II).~\footnote{This also follows from the
structure of the BRST charge (Appendix B), where the $U(1)$ ghost field
$\tilde{c}$ multiplies the $U(1)$ current $J(z)$.}
With conformal dimensions $h\left[e^{\tilde{\s}/2}\right]=-\fracmm{1}{8}$
and $h\left[t^3\right]=+\fracmm{1}{8}$, the conformal
dimension of the ground state in the $\ch^{\rm NS}_{+-+-}$ sector adds up to 0,
as given in Table~II.
Note that the local $U(1)$ charge~$e$ is not defined here since the
$U(1)$ current $J(z)$ is half-integral moded.
This kills the spectral flow and also
explains the blank slots in the last line of Table~II.
BRST invariance will further constrain this state~(sect.~7).

A different method
to calculate conformal dimensions~$h$ (or critical intercepts) associated with
these ground states, consists of collecting the corresponding contributions to
the intercept from the
periodic (P) or anti-periodic (A) world-sheet bosons and fermions in
each sector separately. The standard results~\cite{gsw} are displayed in
Table~III. Our results for the $N{=}2$ string critical intercepts $(-h)$
are given in the last row of Table~II.

\noindent{\sf Table III.} Contributions to (minus) the intercept of matter and
ghost fields for periodic (P or R) and antiperiodic (A or NS)
boundary conditions.
\vglue.1in
\begin{tabular}{c|cc|ccc} \hline
sector & ${\rm complex}\atop{\rm boson}$ & ${\rm complex}\atop{\rm fermion}$ &
$(b,c)$ & $(\tilde{b},\tilde{c})$ & $(\b^i,\g^i)$ \\ \hline
P or R & $-\fracmm{1}{12}$ & $+\fracmm{1}{12}$ & $+\fracmm{1}{12}$ &
$+\fracmm{1}{12}$ & $-\fracmm{1}{12}$ \\
A or NS & $+\fracmm{1}{24}$ & $-\fracmm{1}{24}$ & $-\fracmm{1}{24}$ &
$-\fracmm{1}{24}$ & $+\fracmm{1}{24}$ \\ \hline
\end{tabular}
\vglue.2in

Completing the list of the candidate ground states for all the
twisted sectors listed in Table~II, we find (multiple sign choices
are correlated)
\vglue.1in
\noindent\begin{tabular}{rrr}
$\ch^{\rm NS,R}_{++++}~:$ & \qquad $\ket{k;-1,-1}$ & \qquad
$S^{2,\ua\da}S^{3,\ua\da}\ket{k;-\ha,-\ha}$ \\
$\ch^{\rm NS,R}_{++--}~:$ & \qquad $t^1S^{1,\pm}\ket{k;-1,-1}$ & \qquad
$t^1S^{0,\pm}\ket{k;-\ha,-\ha}$ \\
$\ch^{\rm NS,R}_{--++}~:$ & \qquad $t^0S^{0,\pm}\ket{k;-1,-1}$ & \qquad
$t^0S^{1,\pm}\ket{k;-\ha,-\ha}$ \\
$\ch^{\rm NS,R}_{----}~:$ & \qquad $t^0t^1S^{0,\pm}S^{1,\mp}\ket{k;-1,-1}$ &
\qquad $t^0 t^1 \ket{k;-\ha,-\ha}$ \\
\end{tabular}
\vglue.1in
\noindent whereas
\vglue.1in
\noindent\begin{tabular}{rrr}
$\ch^{\rm NS,R}_{+-+-}~:$ &
\qquad $t^3 S^{3,\ua\da}e^{\tilde{\s}/2}\ket{k;-1,-\ha}$ &
\qquad $t^3 S^{2,\ua\da}e^{\tilde{\s}/2}\ket{k;-\ha,-1}$ \\
$\ch^{\rm NS,R}_{+--+}~:$ &
\qquad $t^{03}t^{12}S^{03}S^{12}e^{\tilde{\s}/2}\ket{k;-1,-\ha}$ &
\qquad $t^{03}t^{12}S^{02}S^{13}e^{\tilde{\s}/2}\ket{k;-\ha,-1}$ \\
\end{tabular}
\vglue.1in
\noindent
and quite similar for $\ch^{\rm NS,R}_{-+-+}$ and $\ch^{\rm NS,R}_{-++-}$.
In all cases, the momenta will have to be adjusted for BRST invariance.

To compute the values of $h$ for zero momentum in all sectors,
it is useful to collect
$$\eqalign{
h\left[\phantom{\ha}\!\!\!\ket{0;-1,-1}\right]=\phantom{ - }0~,&\quad\qquad
h\left[\ket{0;-\ha,-\ha}\right]=-\frac14~,\cr
h\left[\ket{0;-1,-\ha}\right]=-\frac18~,&\quad\qquad
h\left[\ket{0;-\ha,-1}\right]=-\frac18~.\cr}\eqno(4.23)$$
With the values given above and in Table~I we obtain massless states
in all sectors but (4),(6) and~(8) of Table~II.
The local $U(1)$ charges are only defined in the first eight sectors.
Neutrality is impossible in sectors (3)--(6) of Table~II.

The spacetime interpretation of the ${\bf Z}_2$ twists considered so far is
quite clear.
Depending on how many of the three twists ${\bf Z}_2$, ${\bf Z'}_2$ and
${\bf Z''}_2$ for ~$Z$ we are permitting, the target space will be an orbifold
$$\cb\,=\;{\bf C}^{1,1}/{\bf Z}_2 \quad{\rm or}\quad{\bf C}^{1,1}/{\bf Z}_2^2
\quad{\rm or}\quad{\bf C}^{1,1}/{\bf Z}_2^3~,\eqno(4.24)$$
which represents a cone or a halfspace, a quarter- or an eighth-space
in ${\bf C}^{1,1}$, respectively.
The first eight sectors of Table~II contain spacetime bosons, whereas the
ground states of sectors (9)--(16) are good candidates for spacetime fermions.
\vglue.2in

\section{Spectral flow}

We are now going to consider more general monodromies parametrized by arbitrary
phases,
$$\eqalign{
Z^{\m\pm}(\p)\ =&\ +e^{\pm 2\p ia^{\m}}\ Z^{\m\pm}(0)~,\cr
\J^{\m\pm}(\p)\ =&\ -e^{\pm 2\p ia^{\m}}\ e^{\pm 2\p i\n}\ \J^{\m\pm}(0)~,\cr
\c_\a^\pm(\p)\ =&\ -e^{\pm 2\p i\n}\ \c_\a^\pm(0)~,\cr}\eqno(5.1)$$
where the angles $a^\m$ and $\n$ are valued in $S^1={\bf R/Z}$,
and the signs are correlated.
The special values $a^{\m}=0$ and  $a^{\m}=\ha$ correspond to the cases
considered in sect.~3, namely sectors (1)--(8).
Similarly, the values $\n=0$ and $\n=\ha$ correspond to the
NS- and R- sectors, respectively, in our notation.
Stated differently, the complex vectors $Z^\pm$ and $\J^\pm$ are subject
to $U(1)\otimes U(1)$ boundary conditions, with a relative angle~$\n$ between
$Z$ and~$\J$.
We call these monodromies {\it complex elliptic}, for reasons to be explained
in the following section.
Clearly, a `good' target space arises only in case $a^\m$ are rational,
which leads to $\cb={\bf C}^{1,1}/({\bf Z}_n\otimes{\bf Z}_m)$, i.e. a product
of cones.

It is worthy to notice here that the ${\bf Z''}_2$-twisted (complex elliptic)
boundary conditions of the type
$$\eqalign{
Z^{\m\pm}(\p)\ =&\ +e^{\pm 2\p ia^{\m}}\ Z^{\m\mp}(0)~,\cr
\J^{\m\pm}(\p)\ =&\ -e^{\pm 2\p ia^{\m}}\ e^{\pm 2\p i\n}\ \J^{\m\mp}(0)~, \cr
\c_\a^\pm(\p)\ =&\ -e^{\pm 2\p i\n}\ \c_\a^\mp(0)~,\cr}\eqno(5.2)$$
do not give anything new, since the phases in eq.~(5.2) can easily be
removed up to signs by rescaling the fields~\cite{mm}, leaving just the
sectors (9)--(16) of Table~II.

The angle $\n$ in eq.~(5.1) parametrizes rotations between the NS- and
R- sectors, and it can be identified with the parameter of the {\it spectral
flow} in the $N{=}2$ superconformal algebra \cite{ov,lvw}. To understand this
specific feature of the $N{=}2$ string,
one has to investigate the $N{=}2$ supermoduli space. It is not hard to see
that different choices of $\n$ in eq.~(5.1) are simply related by shifts in the
complex moduli of the $U(1)$ gauge field~$A$ \cite{ov}.
Since we are instructed to finally integrate over all moduli, any amplitude has
to be averaged over $\n$ eventually.
It seems to follow that monodromy sectors related by spectral flow should be
identified, as happens in {\it globally} $N{=}2$ superconformal field theory.
This is not really true, for the following reason.
For a given $n$-string amplitude,
the $U(1)$ moduli are represented by non-trivial Wilson lines.
Hence, we may shift the fermionic monodromies for each homology cycle
individually, in particular for the cycles around the punctures.
However, there are $n{-}1$ {\it independent} homology cycles for a tree-level
$n$-string amplitude.
Consequently, exactly {\it one} of the fermionic boundary conditions of the
different asymptotic string states cannot be altered by the spectral flow.
More specifically, we can rotate all {\it but one} $\n$~parameters to zero.
Of course, the amplitude will have to vanish unless the `last' $\n$ was zero
from the beginning.
However, the vanishing of an amplitude does not mean that any of its external
states does not exist.
In the example of a two-point function, we should indeed identify the NS-NS
with the R-R propagator (and all interpolations); however, the NS-R correlation
function (and all others with a relative twist) are genuinely different,
albeit zero.
In summary, the spectral flow does relate some string amplitudes, but does
{\it not} identify NS with R states physically.
Instead, we have a one-parameter family of distinct states labelled by
$\n\in[0,1]$.

To calculate the ground state dimensions for the generalized boundary
conditions in eq.~(5.1),
one needs to sum up the relevant contributions from the $N{=}2$ string fields.
The basic formulae for the vacuum energy of a twisted complex
boson and a twisted complex fermion are \cite{dhvw,har}
$$\eqalign{
h_{\rm B}\ &=\ 2\!\sum_{n\in{\bf Z}+a}\!\fracmm{n}{2}\ =\
\left.\sum_{n\in{\bf Z}}(n{+}a)^{-s}\right|_{s=-1}=\ \zeta(-1,a)\
=\ -\fracm{1}{12} +\ha a(1{-}a)\ =:\ f(a)~,\cr
h_{\rm F}\ &=\ +\fracm{1}{12} -\ha(a+\n+\ha)(1-a-\n-\ha)\ =\ -f(a+\n+\ha)~,\cr}
\eqno(5.3)$$
where $a\in[0,1]$ in the first line and $a+\n+\ha\in[0,1]$ in the second
one.~\footnote{Note that this expression is not periodic.}
This result forces the conformal dimensions of general complex
twist fields and spin fields to be equal to $h[t]=\frac12 a(1{-}a)$ and
$h[S]=\frac12(a+\n)^2$, respectively.
In adding the individual contributions of the different fields,
care has to be exercised in order to make sure that the periodic argument of
the function~$f$ is always taken from the interval~$[0,1]$, where eq.~(5.3)
applies.
In general, one has to create a list of case distinctions~\cite{mm}.
The results of counting are summarized in Table~IV.

\newpage

\noindent{\sf Table~IV}. Contributions to the vacuum energy. The left column
corresponds
to the generalized boundary conditions in eq.~(5.1). The right column gives
the result for the discrete ${\bf Z''}_2$-twisted cases (9)--(16) of Table~II.
$\n\in[{-}\ha,\ha]$ and $a^\m\in[{-}\ha{-}\n,\ha{-}\n]$.
\vglue.1in
\begin{tabular}{c|c|c}
\hline
 & $a^0 \qquad\, a^1$ & $Z^\pm{\to}Z^\mp~{\rm and}~\j^\pm{\to}\j^\mp$ \\
twists & ${\pm\pm\atop\pm\pm}\qquad{\pm\pm\atop\mp\mp}$
 & $\pm\mp\pm\mp\atop\pm\mp\mp\pm$ \\
\hline
$Z$ & $f(a^0)+f(a^1)$ & $f(0)+f(\ha)$ \\
$\j$ & $-f(a^0+\n+\ha)-f(a^1+\n+\ha)$ & $-f(0)-f(\ha)$ \\
$b,c$ & $-f(0)$ & $-f(0)$ \\
$\b,\g$ & $2f(\n+\ha)$ & $f(0)+f(\ha)$ \\
$\tilde{b},\tilde{c}$ & $-f(0)$ & $-f(\ha)$ \\
\hline
$h{=}h^0{+}h^1$ & $h^\m=\cases{a^\m(\n+\ha) & for $a^\m\in[0,\ha{-}\n]$\cr
a^\m(\n-\ha) & for $a^\m\in[{-}\ha{-}\n,0]$\cr}$ & $0$ \\
\hline
\end{tabular}
\vglue.2in

\noindent
The terms quadratic in $a^\m$ cancel among bosons and fermions due to
$$f(a)-f(a+\n+\ha)\ =\ a(\n+\ha)+(\n^2-\frac14)~.\eqno(5.4)$$
The $U(1)_{\rm gauged}$ charge for the ground states in question is given by
$|e|=(|a^0|+|a^1|)$, where it is defined.
It follows that, for a given $\n$, $h$ is a symmetric, periodic and piece-wise
linear function of $a^0$ and $a^1$, whose derivatives jump on the lines
$a^\m=0$ as well as $a^\m=\ha{-}\n$ and $a^\m={-}\ha{-}\n$.
Moreover, $h$ is positive everywhere except at $a^0=a^1=0$ and for $\n=\pm\ha$
where it vanishes identically.

The ground state energy equally follows from computing the $N{=}2$
superconformal algebra in terms of the $N{=}2$ string BRST currents. The
relevant central extension terms $A(m)$ in the Virasoro subalgebra,
having the form
$$\[L^{\rm tot}_m,L^{\rm tot}_n\]\ =\ (m-n)L^{\rm tot}_{m+n} +A(m)\d_{m+n,0}~,
\eqno(5.5)$$
are at most {\it cubic} in $m$.~\footnote{For instance, $A(m)
=\frac{1}{12}(1-3\tilde{Q}^2)m^3+2h_{\rm F}m$ for an anticommuting
$(b,c)$-type system with background charge $\tilde{Q}$.} The
total $m^3$-contribution to $A(m)$, being independent of the twisting phases,
vanishes in the critical dimension. The coefficients of the terms linear in
$2m$ sum to minus the critical intercept; they are collected in Table~III.
See refs.~\cite{mm,thesis} for the explicit calculations in this approach.

The conclusions we can draw for the $N{=}2$ string from the spectral flow
analysis are the following. First, the corresponding R- and NS-sectors of
Table~II are {\it not} to be identified although they are related by
spectral flow. This differs from one of the conclusions of ref.~\cite{ov}.
In fact, one has a one-parameter family of distinct states here.
Second, the conformal dimension~$h$ of the ground state
{\it changes} under the spectral flow unless $a^0=a^1=0$.
If we accept only {\it massless gauge-singlet} states as
physical~\cite{bien,bh},~\footnote{This is naively `obvious' in the `double'
light-cone gauge for the $N{=}2$ string, where all the
excitations of $Z$'s and $\J$'s disappear.} the two angles
$a^\m$ in eq.~(5.1) are required to vanish.
Although the Ramond ground state is always massless, we would generally
consider it to be unphysical, from a continuity argument.
Hence, only states without $U(1)\otimes U(1)$ twist and, in particular,
without ${\bf Z}_2$ or ${\bf Z'}_2$ twist survive, leaving only the
(Ooguri-Vafa) states, i.e. sectors (1) and~(2),
from the first 8~sectors of Table~II.
Third, this discussion does not apply to the ${\bf Z''}_2$-twisted states,
where all sectors (9)--(16) remain.
Their invariance under spectral flow is in line with the twisting of the gauge
field, which kills all $U(1)$ moduli. Each of these discrete sectors may be
generated from, say, sector~(9) by the action of ${\bf Z}_2$, ${\bf Z'}_2$
and ${\bf Z'''}_2$ twists in~(3.2).

All this
gives reasons to identify the ground states in the untwisted (Ooguri-Vafa)
sectors (1) and~(2) as physical `spacetime' {\it bosons},
whereas the ground states in the (Mathur-Mukhi) sectors (9)--(16)
should be physical `spacetime' {\it fermions}.
Demanding the ground states of all occurring sectors to be {\it massless},
only ${\bf Z''}_2$ and ${\bf Z'''}_2$ twists are permitted,
selecting sectors (1),(2) and (9),(10) at most.
Since each sector contributes only one or two physical degrees of freedom,
${\bf Z}_2$ twists leave room for a $2_{\rm B}\oplus2_{\rm F}$ `spacetime'
supermultiplet in the complex half-space $\cb={\bf C}^{1,1}/{\bf Z''}_2$.
These observations partially support the idea of `space-time'
extended supersymmetry in $N{=}2$ string theory, put forward in
ref.~\cite{siegel}.
However, nontrivial target space topology seems to be needed to realize it.
In principle, the spectral flow parameter~$\n$ may take values different from
$0$ or~$\ha$. The corresponding vertex operators, however, have little chance
of leading to a local operator algebra (see sect.~7).
\vglue.2in

\section{General monodromy conditions}

We are now in a position to discuss the most general monodromy conditions for
the $N{=}2$ string. Let us concentrate on the $Z$ fields, and arrange them
into a complex doublet as
$$ Z=\left(\begin{array}{c} Z^{0+} \\ Z^{1+}\end{array} \right)~,\qquad
Z^{\dg}=\left( Z^{0-}, Z^{1-} \right)~,\eqno(6.1)$$
where $Z^0=Z^{02}+iZ^{03}~,~~Z^1=Z^{12}+iZ^{13}$. The fields $\j$ and
$\c_{\a}$ can be treated similarly. The fields $Z$ enter the $N{=}2$ string
action via the kinetic term, $\pa_\pm Z^{\dg}\cdot\h\cdot\pa_\mp Z$,
where $\h={\rm diag}(-,+)$ is the two-dimensional complex target space metric.
The kinetic term is obviously invariant under the unitary transformations
$$Z\to UZ~,\qquad Z^{\dg}\to  Z^{\dg}U^{\dg}~,
\qquad{\rm with}\quad U^{\dg}\h U=\h~.\eqno(6.2) $$
As remarked earlier, these $U(1,1)$ transformations, together with similar
compensating `rotations' of $\J$ and $\c$, actually constitute the global
continuous symmetry group of the {\it full} action~(3.1).

We shall now consider arbitrary $U(1,1)$ transformations~(6.2) as possible
monodromies ($z\to e^{2\pi i}z$).
The determinantal phase factor may be split off trivially.
This is just the $U(1)$ factor corresponding to $a^0+a^1$, which was ruled
out in the last section.
The remaining $SU(1,1)$ subgroup can be realized as
$$ U =\left(\begin{array}{cc} \a & \b \\ \b^* & \a^* \end{array} \right)~,
\qquad \abs{\a}^2-\abs{\b}^2=1~.\eqno(6.3)$$
The corresponding $su(1,1)$ Lie algebra generators satisfy the relations
$$\[L_1,L_2\]=-L_3~,\qquad \[L_2,L_3\]=+L_1~,\qquad \[L_3,L_1\]=+L_2~,
\eqno(6.4)$$
where $L_1$ and $L_2$ are non-compact and hermitian, and $L_3$ is compact and
anti-hermitian. In our basis (6.1) $L_3$ turns out to be diagonal:
$$L_1=\ha\left(\begin{array}{cc} 0 & 1 \\ 1 & 0 \end{array} \right)~, \qquad
L_2=\ha\left(\begin{array}{cc} 0 & i \\ -i & 0 \end{array} \right)~, \qquad
L_3=\ha\left(\begin{array}{cc} i & 0 \\ 0 & -i \end{array} \right)~.
\eqno(6.5)$$

Of course, the matrix representation depends on the choice of basis.
For a different choice, the form of the hermitian metric $\h$ changes, and it
dictates the structure of the $SU(1,1)$ monodromy matrix via eq.~(6.2).
In general, the only restrictions on the metric are
$\h^{\dg}=\h$ and $\det\h=-1$, so that
$$\h\ =\ \left(\begin{array}{cc}
\h_0+\h_3 & \h_1-i\h_2 \\ \h_1+i\h_2 & \h_0-\h_3 \end{array} \right)\
\equiv\ \h_{\m}\s^{\m}~, \qquad \h_{\m}\h^{\m}=1~,\eqno(6.6)$$
where $\s^{\m}=({\bf 1},\vec{\s})$, and $\vec{\s}=(\s_1,\s_2,\s_3)$ are Pauli
matrices. Each value for $\h$ determines a basis for the target
space components of the fields.
Different bases are related by linear field redefinitions,
$$Z\to \tilde{Z}=MZ~,\qquad Z^{\dg}\to \tilde{Z}^{\dg}=Z^{\dg}M^{\dg}~,
\eqno(6.7)$$
forming the linear matrix group $GL(2,{\bf C})$.
The redefined field $\tilde{Z}$ has the monodromy
$$\tilde{Z}\equiv MZ\ \to\ MUZ=(MUM^{-1})MZ=(MUM^{-1})\tilde{Z}
\equiv\tilde{U}\tilde{Z}~, \eqno(6.8)$$
so that eq.~(6.7) induces
$$ U\ \to\ \tilde{U}\ =\ MUM^{-1}~.\eqno(6.9)$$
At the same time, the action is unchanged,
$$\tilde{Z}^{\dg}\tilde\h\tilde{Z}\ =\ Z^{\dg}\h Z~,\eqno(6.10)$$
so that the new metric takes the form
$$\tilde{\h}\ =\ M^{-1\dg}\h M^{-1}~.\eqno(6.11)$$
Any two metrics~(6.6) are related this way.
Apparently, the determinantal factor of $M$ merely leads to a complex
rescaling, which is rather trivial. We therefore take $M\in SL(2,{\bf C})$.
When $M\in SU(1,1)$, eq.~(6.11) becomes $\tilde{\h}=\h$, and, hence,
$\tilde{U}$ is conjugate to $U$ in $U(1,1)$.
However, when $M\notin SU(1,1)$, one gets $\tilde{\h}\neq \h$,
which implies a change in the metric eigenvalues.

Given a fixed basis and metric $\h$, we must consider two monodromies as
equivalent if they are related by a global symmetry transformation
$M\in U(1,1)$. Therefore, if we are interested in {\it all} inequivalent
monodromies~$U$ of the $N{=}2$ string, we need to consider all the
{\it conjugacy
classes} of $U(1,1)=\left[ U(1)\otimes SU(1,1)\right]/{\bf Z}_2$.~\footnote{
plus the ${\bf Z''}_2$ twist.}
These are labelled uniquely\footnote{except if $|\tr\,U|=2$}
by the determinant $\det\,U$, corresponding to the $U(1)$ factor,\footnote{
We can actually take $(\det\,U)^2$ due to the sign ambiguity
from ${\bf Z}_2$.}
and the normalized trace
$(\det\,U)^{-1/2}\,\tr\,U$, corresponding to the trace in $SU(1,1)$.
The absolute value of the
normalized trace  can be either less than $2$, equal to $2$, or greater than
$2$, which distinguishes the so-called {\it elliptic}, {\it parabolic} and
{\it hyperbolic} cases, respectively.

For a generic $SU(1,1)$ monodromy of the form
$$\left(\begin{array}{c} Z^0 \\ Z^1\end{array} \right)(\s+\p)\ =\
\left(\begin{array}{cc} \a & \b \\ \b^* & \a^* \end{array} \right)
\left(\begin{array}{c} Z^0 \\ Z^1\end{array} \right)(\s)~,\eqno(6.12)$$
substituting the mode expansion
$$\left(\begin{array}{c} Z^0 \\ Z^1\end{array}\right)(\s)\ =\ \sum_n e^{2in\s}
K(\s)\left(\begin{array}{c} a^0_n \\ a^1_n \end{array} \right)\eqno(6.13)$$
yields the equation
$$K(\s+\p)\ =\ \left(\begin{array}{cc}\a & \b \\ \b^* &\a^* \end{array}\right)
K(\s)~.\eqno(6.14)$$
With $K(0)={\bf 1}$ we have $U=K(\p)$.
Introducing the $su(1,1)$ generators as
$$\left(\begin{array}{cc} \a & \b \\ \b^* & \a^* \end{array} \right)\ =\
\exp\left[\sum^3_{j=1}\q_jL_j\right]~,\eqno(6.15)$$
we find the solution in the form
$$K(\s)\ =\ \exp\left[\fracmm{\s}{\p}\sum^3_{j=1}\q_jL_j\right]~.\eqno(6.16)$$

Let us be more precise for the basis (6.1), i.e. when $\h=diag(-,+)$.
If just the compact generator appears, $\q_j=2\p\q\d_{j3}$, one gets
$$K_{\rm ell}(\s)\ =\ \left(\begin{array}{cc} e^{i\q\s} & 0 \\ 0 &  e^{-i\q\s}
\end{array} \right)~,\eqno(6.17)$$
which obviously corresponds to the elliptic case ($|\tr\,U|{<}2$).
Indeed, setting $\q=a^0{-}a^1$ and multiplying $U_{\rm ell}=K_{\rm ell}(\p)$
with
the $U(1)$ phase $e^{i\p(a^0+a^1)}$ just reproduce the
complex elliptic monodromy of eq.~(5.1).
As was already mentioned, the target space is a tensor product of two cones
for the case of rational angles.

Turning on a non-compact generator, e.g. $\q_j=2\p\q\d_{j1}$, leads to the
prime example of a hyperbolic class,
$$K_{\rm hyp}(\s)\ =\ \left(\begin{array}{cc}\cosh\q\s &\sinh\q\s \\ \sinh\q\s
& \cosh\q\s \end{array} \right)~.\eqno(6.18)$$
Here, we find $|\tr\,U|{>}2$, and $U\in SO(1,1)$ is
the `real' Lorentz symmetry already discussed in sect.~3.
The mode expansion becomes complex and matrix-valued,
$$\left(\begin{array}{c}Z^0\\ Z^1\end{array}\right)(\s)\ =\ \sum_n\ \exp\left[
2i\left(\begin{array}{cc} n & -i\q/2 \\ -i\q/2 &  n \end{array} \right)\s
\right]\left(\begin{array}{c} a^0_n \\ a^1_n \end{array} \right)~.\eqno(6.19)
$$
The alternative is to switch to a basis where $L_1$ is diagonal.
This is achieved by taking
$$M=\fracmm{1}{\sqrt{2}}\left(\begin{array}{cc}1&1\\1&{-}1\end{array}\right)~,
\quad \tilde{L}_1=M\,L_1\,M^{-1}=
\ha\left(\begin{array}{cc}1&0\\0&{-}1\end{array}\right)~,
\quad \tilde{\h}= \left(\begin{array}{cc}0&{-}1\\{-}1&0\end{array}\right)~,
\eqno(6.20)$$
which simply brings us to the familiar light-cone basis,
$\tilde{Z}=\fracmm{1}{\sqrt{2}}\left({Z^\ua\atop Z^\da}\right)$.
Here, of course, we just get
$$\tilde{K}_{\rm hyp}(\s)\ =\ \left(\begin{array}{cc} e^{\q\s}&0\\0&e^{-\q\s}
\end{array}\right)~,\eqno(6.21)$$
but the plane wave modes remain complex.
The target space interpretation is also more transparent in this basis.
The identifications
$$Z^\ua\cong e^{\q\p}Z^\ua~, \qquad Z^\da\cong e^{-\q\p}Z^\da~\eqno(6.22)$$
create two similar tori (with real modulus~$\q$).
Hence, the background~$\cb$ is just their tensor product.
One should remark that in this situation an additional ${\bf Z''}_2$ twist
does not allow us to set $\q{=}0$ by rescaling $Z$, as was the case in
eq.~(5.2). Here, it leads to yet another target space.

Rather unusual is the parabolic situation ($|\tr\,U|=2$) which is generated
by either one of the nilpotent combinations
$$N_\pm\ =\ L_3\pm L_2\ =\ \ha\left(\begin{array}{cc} i & \pm i \\ \mp i & -i
\end{array} \right)\eqno(6.23)$$
and yields
$$K_{\rm par}(\s)\ =\ \left(\begin{array}{cc}
1+i\q\s&\pm i\q\s\\ \mp i\q\s&1-i\q\s \end{array} \right)~.\eqno(6.24)$$
In the light-cone basis, we have
$$\tilde{N}_+\ =\ \left(\begin{array}{cc}0&0\\i&0\end{array} \right)~,\qquad
\tilde{N}_-\ =\ \left(\begin{array}{cc}0&i\\0&0\end{array} \right)~,$$
$$\tilde{K}_+\ =\ \left(\begin{array}{cc}1&0\\2i\q\s&1\end{array} \right)~,
\qquad\tilde{K}_-\ =\ \left(\begin{array}{cc}1&2i\q\s\\0&1\end{array}\right)~.
\eqno(6.25)
$$
The background $\cb$ in this case is remarkable.
{}From $\tilde{K}_-$, for example, we read off the identifications
$$Z^\ua\cong Z^\ua+2\p\q(iZ^\da)~, \qquad (iZ^\da)\cong(iZ^\da)~,\eqno(6.26)$$
which is topologically a complex cone.
The entire $iZ^\da{=}0$ axis has to be identified to a single point through
which pass (almost) all straight lines.~\footnote{
If this is not done, one ends up with a non-Haussdorff space.}
The metric, $\tilde{\h}=-\s_1$, on this cone degenerates for two different
real sections: either $Z^{\ua\da-}={\ua\da}Z^{\ua\da+}$ or
$Z^{\da\pm}=\pm iZ^{\ua\pm}$ leads to a vanishing of
$|Z|^2=Z^{\ua-}Z^{\da+}+Z^{\da-}Z^{\ua+}$.
This leaves room for null planes, just like in ${\bf C}^{1,1}$.
It is also exceptional that $|\tr\,U|=2$ labels not one but {\it three}
conjugacy classes: the identity class as well as the two parabolic ones
generated by $N_+$ and $N_-$.

One should add that any $SU(1,1)$ monodromy matrix~(6.15) can be factorized
into a product of three matrices, with each being in one of the standard forms
displayed above. However, there is always a basis in which it just takes
the standard form corresponding to its (elliptic, hyperbolic or parabolic)
nature.

The ground state energies calculated for the complex elliptic case in sect.~5
are all based on
eq.~(5.3). This equation needs to be generalized to the other cases
(hyperbolic or parabolic).
The complex elliptic case corresponds to
$$K(\s)\ =\ \left(\begin{array}{cc} e^{2ia^0\s} & 0 \\ 0 &  e^{2ia^1\s}
\end{array} \right)\ =\ \exp\left[2i\left(\begin{array}{cc} a^0 & 0 \\ 0 & a^1
\end{array} \right)\s\right]\ =:\ \exp\left[2i\s\ca_{\rm real}\right]~,
\eqno(6.27)$$
and leads to the value  $f(a^0)+f(a^1)$ in Table~III,
with $f(a)=-\fracmm{1}{12}+\fracmm{1}{2}a(1{-}a)$.
The natural attempt is to view the result as arising from the trace
$$2 f(\ca)\ =\ -\frac16+\frac12\tr\left[\ca({\bf1}-\ca)\right]~,\eqno(6.28)$$
with the matrix $\ca=\ca_{\rm real}$ of real angles taken from a `fundamental
domain' since, again, this expression is not periodic.
For a general $U\in U(1,1)$, one should factor off the $U(1)$ phase
$e^{i\q_0/2}$ and use the $su(1,1)$ algebra parametrization (6.15).
Inserting
$$\ca(\q)\ =\ \fracmm{1}{2\p i}\left[\q_0\fracmm{i}{2}{\bf 1}+
\sum_{j=1}^3 \q_j L_j\right]~\eqno(6.29)$$
into the {\it basis-independent} formula~(6.28) we arrive at
$$2f(\q)\ =\ -\fracmm{1}{6}+\fracmm{1}{16\p^2}\q_0\left(4\p{-}\q_0\right)+
\fracmm{1}{16\p^2}\left(\q_1^2+\q_2^2-\q_3^2\right)~, \eqno(6.30)$$
for $\q_0\in[0,4\p]$ and
$\q_3\in\left[|\q_0{-}2\p|{-}2\p,{-}|\q_0{-}2\p|{+}2\p\right]$.
These restrictions arise from $a^\m\in[0,1]$ and the relations
$\q_0=2\p(a^0{+}a^1)$ and $\q_3=2\p(a^0{-}a^1)$.
The quadratic form in~$\q_j$ is nothing but the invariant length-squared
of a boost or rotation vector associated to the action of $so(2,1)\cong
su(1,1)$
in (2+1)-dimensional Minkowski space.
Time-, space- and light-like vectors are seen to belong to elliptic, hyperbolic
and parabolic conjugacy classes, respectively.
Remarkably, the ground state energy takes the untwisted value only for
$\q_0=0~{\rm mod}~4\p$ and a {\it light-like} vector~$\{\q_i\}$,
i.e. in the identity and parabolic classes, with possible ${\bf Z''}_2$ twist.
A direct calculation of the $L_0$ eigenvalue of the ground state confirms this.
We already know from Table~IV that {\it elliptic} $SU(1,1)$ twists change the
conformal dimension of the ground state, since for $\q_0{=}\q_1{=}\q_2{=}0$
and $|\n|\leq\ha$ we found
$$h\ =\ \cases{
|\q_3|/4\p & \quad if $|\q_3|\leq4\p(\n{+}\ha)$~, \cr
\n{+}\ha & \quad if $|\q_3|\geq4\p(\n{+}\ha)$~. \cr}
\eqno(6.31)$$
For hyperbolic twists, however, the `angles' $\q_1$ and $\q_2$ are no longer
periodic, and the naive summation of contributions to the ground state energy
based on eq.~(6.30) gives a vanishing result.
Looking back to the lesson learned in sect.~5, we should nevertheless expect
a vanishing ground state energy only for $\n{=}\ha$ and for at most
${\bf Z''}_2$ and/or parabolic twists.
\vglue.2in

\section{BRST cohomology and interactions}

We have assumed in sect.~4 that the spin fields $S$ ($S^i$ or $S^{\m}$),
as well as the twist fields $t$ ($t^i$ or $t^{\m}$)
are all primary and have conformal dimensions
$h{=}\fracm{1}{8}$. This does not mean, however, that they are superconformal
primaries. They are {\it not}, in fact, as follows from examining the OPEs of
these fields with the $N{=}2$ superconformal algebra generators
$G^\pm=\j^\pm\sdot\pa Z^\mp$ and $J=\fracm{1}{4} \j^-\sdot\j^+$.
In particular, the OPE $G^\pm(z)(St)(w)$ gives rise to the new field
$(\G^\pm S)\cdot\tilde{t}^\mp$ of dimensions $h=\fracmm{3}{4}$ and
$\Bar{h}=\fracmm{1}{4}$, where `spacetime' gamma matrices
$\G^{\m\pm}=\G^{\m2}\pm i\G^{\m3}$ have been introduced.~\footnote{
According to eqs.~(4.19) and
(4.20), the twist field $\tilde{t}$ has dimensions $h=\fracmm{5}{8}$ and
$\bar{h}=\fracmm{1}{8}$.}
Depending on the spin field, the gamma matrices act either on the $U(1)$
index as $(\G^{\m i})^{\pm,\pm}$, or on the $O(1,1)$ index as
$(\G^{\m i})^{\ua\da,\ua\da}$.

Our task now is to consider an action of the BRST charge in the form
\cite{li}
$$Q\ =\ Q_1 + Q_2 + Q_3~,\qquad Q_i\ =\ \oint j_i(z)~,\qquad \oint\equiv
\oint_0\fracmm{dz}{2\p i}~,$$
$$\eqalign{
j_1\ =&\ cT_{\rm tot} +bc\pa c~,\cr
j_2\ =&\ \ha\left( \g^+G^-+\g^-G^+ + 2\tilde{c}J\right)~,\cr
j_3\ =&\ -b\g^+\g^--\tilde{b}\left(\g^+\pa\g^--\g^-\pa\g^+
\right)+\fracm14\tilde{c}\left(\b^+\g^--\b^-\g^+\right)~,\cr}\eqno(7.1)$$
on the candidate ground states or the corresponding vertex operators creating
these states.

First, the BRST operator of eq.~(7.1) annihilates the (untwisted) NS ground
state $\ket{k}^{\rm NS}_{++++}=\ket{k;-1,-1}$ when $k^+\sdot k^-=0$,
The same is also true for the (untwisted) R~ground states
$\ket{k}^{\rm R}_{++++}=S^{2,\ua\da}S^{3,\ua\da}\ket{k;{-}\ha,{-}\ha}$,
provided that $k^{\ua2}=0=k^{\ua3}$ or $k^{\da2}=0=k^{\da3}$,
for the upper or lower choice of sign, respectively.
These momentum constraints imply $k^+\sdot k^-=0$ but are more stringent.

Second, let us investigate sectors (3) and~(4) of Table~II, where the
combined twist $(St)^1$ or $S^0t^1$ acting on $\ket{k}^{\rm NS}_{++++}$ yields
$\ket{k}^{\rm NS}_{++--}$ or $\ket{k}^{\rm R}_{++--}$, respectively.
To this end, consider the OPEs of the BRST currents $j_i(z)$ with the
NS vertex operator of eq.~(4.21), expressed in complex notation,
$$V_{(4)}(w)\ =\ \bar{u}\,c\,\d(\g^+)\,\d(\g^-)\,(St)^1\,
e^{\frac{i}2(k^+\cdot Z^-+k^-\cdot Z^+)}(w)~,\eqno(7.2)$$
which creates the
$\ch^{\rm NS}_{++--}$ ground state having the spinor wave function $u(k)$.
Using the bosonization formulae~(4.13) and the relevant OPEs
$$\g^\pm(z)\,\d(\g^\pm)(w)\sim(z{-}w)\,\h^\pm(w)~,\qquad
\b^\pm(z)\,\d(\g^\mp)(w)\sim\fracmm{1}{z{-}w}\,\pa\x^\pm~,$$
$$\j^{\m\pm}(z)\,S^\n(w)\sim\fracmm{\h^{\m\n}}{\sqrt{z{-}w}}\G^{\m\pm}S(w)~,
\qquad \pa Z^{\m\pm}(z)\,t^\n(w)\sim\fracmm{\h^{\m\n}}{\sqrt{z{-}w}}
\tilde{t}^{\m\pm}(w)~,\eqno(7.3)$$
we find
$$j_1(z)\ V_{(4)}(w)\ \sim\ \fracmm{h}{z{-}w}(\pa c)V_{(4)}(w)~,\eqno(7.4a)$$
$$\eqalign{j_2(z)\ V_{(4)}(w)\ \sim&\ \fracmm{1/2}{\sqrt{z{-}w}}\bar{u}c\left[
\h^+\d(\g^-)({-}ik^{1+})\G^{1-} + ({+}\leftrightarrow{-})\right]\cr
&\times (St)^1\ e^{\frac{i}2(k^+\cdot Z^-+k^-\cdot Z^+)}(w)
+\fracmm{e/2}{z{-}w}\tilde{c}V_{(4)}(w)~,\cr}\eqno(7.4b)$$
$$j_3(z)\ V_{(4)}(w)\ \sim\ \ha\tilde{c}\fracmm{[1-1]}{z{-}w}V_{(4)}(w)\
\sim O(1)~, \eqno(7.4c)$$
where the last equation (c) follows due to a cancellation among the terms
produced from $\g^-\b^+=2\pa\vf^-$ and $\g^+\b^-=2\pa\vf^+$.
Eq.~(7.4) implies that the ${\bf Z'}_2$-twisted NS ground state is BRST-closed
provided it satisfies the following three conditions:
$$ h=\fracm14+\fracm12k^+\sdot k^-=0~,\qquad
\bar{u}k^{1\pm}\G^{1\mp}=0~,\qquad e\left(=\pm\ha\right)=0~,\eqno(7.5)$$
where $e$ is the {\it local} $U(1)$ charge of the state.
Apparently, $V_{(4)}$ cannot be annihilated by the BRST charge.

Next, we do the same calculation for the $\ch^{\rm R}_{++--}$ sector, where
the relevant vertex operator has the form
$$V_{(3)}(w)\ =\ \bar{u}\,c\,e^{-\vf^+/2}\,e^{-\vf^-/2}\,S^0t^1\,
e^{\frac{i}2(k^+\cdot Z^-+k^-\cdot Z^+)}(w)~.\eqno(7.6)$$
It follows
$$j_1(z)\ V_{(3)}(w)\ \sim\ \fracmm{h}{z{-}w}(\pa c)V_{(3)}(w)~,\eqno(7.7a)$$
$$j_2(z)\ V_{(3)}(w)\ \sim\ \fracmm{e/2}{z{-}w}\tilde{c}V_{(4)}(w)
+~{\rm terms~of~order}~\fracmm{1}{z{-}w}~{\rm and}~\fracmm{1}{\sqrt{z{-}w}}~,
\eqno(7.7b)$$
$$j_3(z)\ V_{(3)}(w)\ \sim\ \ha\tilde{c}\fracmm{[\ha-\ha]}{z{-}w}V_{(4)}(w)\
\sim O(1)~. \eqno(7.7c)$$
Although now $h=\fracm12k^+\sdot k^-$ vanishes for a massless excitation,
again the vertex operator~(7.6) is not BRST invariant, in particular because
$e=\pm\ha\neq0$.

Third, we come to the $(----)$-type boundary conditions.
For the NS vertex operator
$$V_{(8)}(w)\ =\ \bar{u}_0\bar{u}_1\,c\,\d(\g^+)\,\d(\g^-)\,(St)^0(St)^1\,
e^{\frac{i}2(k^+\cdot Z^-+k^-\cdot Z^+)}(w) \eqno(7.8)$$
we get
$$j_1(z)\ V_{(8)}(w)\ \sim\ \fracmm{h}{z{-}w}(\pa c)V_{(8)}(w)~,\eqno(7.9a)$$
$$\eqalign{
j_2(z)\ V_{(8)}(w)\ \sim&\ \fracmm{1/2}{\sqrt{z{-}w}}\bar{u}_0\bar{u}_1\,c
\left[\h^+\d(\g^-)\,i\bigl(k^{0+}\G^{0-}S^0S^1-k^{1+}\G^{1-}S^1S^0\bigr)
\right. \cr
& \left. +\phantom{\h^+}\!\!\!\! (+\leftrightarrow-)\right]\
t^0t^1\,e^{\frac{i}2(k^+\cdot Z^-+k^-\cdot Z^+)}(w)
+\fracmm{e/2}{z{-}w}\tilde{c}V_{(8)}(w)~,\cr}\eqno(7.9b)$$
$$j_3(z)\ V_{(8)}(w)\ \sim O(1)~.\eqno(7.9c)$$
The BRST invariance requires
$$ h=\fracm12+\fracm12k^+\sdot k^-=0~,\qquad
\bar{u}_\m k^{\m\pm}\G^{\m\mp}=0~{\rm(no~sum!)}~,\qquad e=0~.\eqno(7.10)$$
This implies $k^+\sdot k^-=-1$ and $k^{\m\pm}=0$ which cannot be
achieved.

More easily, the $\ch^{\rm R}_{----}$ sector with
$$V_{(7)}(w)\ =\ c\,e^{-\vf^+/2}\,e^{-\vf^-/2}\,t^0t^1\,
e^{\frac{i}2(k^+\cdot Z^-+k^-\cdot Z^+)}(w) \eqno(7.11)$$
yields
$$j_1(z)\ V_{(7)}(w)\ \sim\ \fracmm{h}{z{-}w}(\pa c)V_{(7)}(w)~,\eqno(7.12a)$$
$$\eqalign{j_2(z)\ V_{(7)}(w)\ \sim&\ \fracmm{1/2}{\sqrt{z{-}w}}c\left[
\h^+e^{\vf^+/2}e^{-\vf^-/2}({-}ik^+)\sdot\j^-+({+}\leftrightarrow{-})\right]\cr
& \times t^0t^1\ e^{\frac{i}2(k^+\cdot Z^-+k^-\cdot Z^+)}(w)
 + \fracmm{e/2}{z{-}w}\tilde{c}V_{(7)}(w)~,\cr}\eqno(7.12b)$$
$$j_3(z)\ V_{(7)}(w)\ \sim\ \tilde{c}\fracmm{[\ha-\ha]}{z{-}w}V_{(7)}(w)\
\sim O(1)~, \eqno(7.12c)$$
which is regular if $k^\pm=0$ since $h=\fracm12k^+\sdot k^-$ and $e=0$ here.
We do not consider this as a `spacetime' field degree of freedom.

The last new pattern comes from the discrete $\ch_{+-+-}$ sectors.
It is now more convenient to express the BRST current in the real
component fields, as given in eq.~(B.14) and bosonized in eq.~(4.10).
In particular, looking at $\ch_{+-+-}^{\rm NS}$ with
$$V_{(10)}(w)\ =\ \bar{u}\,c\,e^{\tilde{\s}/2}\,e^{-\vf^2}\,e^{-\vf^3/2}\,
S^3t^3\,e^{i(k^2\sdot Z^2+k^3\sdot Z^3)}(w)~,\eqno(7.13)$$
we obtain
$$j_1(z)\ V_{(10)}(w)\ \sim\ \fracmm{h}{z{-}w}(\pa c)V_{(10)}(w)~,
\eqno(7.14a)$$
$$\eqalign{
j_2(z)\ V_{(10)}(w)\ \sim&\ \fracmm{1/2}{z{-}w}\bar{u}\,ce^{\tilde{\s}/2}
e^{-\vf^2}e^{\vf^3/2}\h^3({-}ik^2_\m)\G^{\m3}S^3t^3\
e^{i(k^2\cdot Z^2+k^3\cdot Z^3)}(w)\cr
&+\fracmm{1}{\sqrt{z{-}w}}~\times({\rm terms~linear~in}~k^3)~,\cr}
\eqno(7.14b)$$
$$j_3(z)\ V_{(10)}(w)\ \sim O(1)~,\eqno(7.14c)$$
where $h=\ha k^i\sdot k^i$.
Notice that the charge-measuring term $\tilde{c}J$ in $j_2$ does not lead
to a singularity. To obtain a vanishing BRST commutator,
we have to request not only $k^i\sdot k^i=0$ and the Dirac equation
$$\eqalign{\bar{u}k^2_\m\G^{\m3}\
&=\ -\ha\bar{u}\left(k^{\ua2}\G^{\da3}+k^{\da2}\G^{\ua3}\right)\cr
&=\ -\ha k^{\ua2}\bar{u}^\da\left(\G^{\da3}\right)^{\ua,\da}
-\ha k^{\da2}\bar{u}^\ua\left(\G^{\ua3}\right)^{\da,\ua}~,\cr}\eqno(7.15)$$
but also set $k^3$ to zero.
The latter is not surprising, since we created the twisted state on the border
of the half-space ${\bf C}^{1,1}/{\bf Z''}_2$ so that its transversal momentum
gets trapped at $k^3=-k^3$.
Furthermore, the Dirac equation~(7.15) reads
$$\bar{u}^\da k^{\ua2}=0~\qquad{\rm and}\qquad\bar{u}^\ua k^{\da2}=0~,
\eqno(7.16)$$
so that one helicity is removed, e.g. by choosing
$k^{\ua2}=0$ and $\bar{u}^\ua=0$, i.e. taking only $S^{3,\ua}$ in eq.~(7.13).
With these requirements, $V_{(10)}$ does create
a BRST-invariant state with light-like momentum $k^{\da2}\neq0$.
Similar calculations for $\ch^{\rm R}_{+-+-}$ as well as the remaining sectors
(11)--(16) arrive at the same conclusion.

The analysis above allows us to make the statement that our candidate ground
states in sectors (1),(2) and (9)--(16) are actually the {\it physical} states,
i.e. they represent the BRST cohomology classes. Indeed, they are BRST-closed,
as we have explicitly showed, while they cannot be BRST-exact
(i.e. of the type $Q\ket{*}$) because there are
no candidates for a star state with the correct ghost and picture
numbers and conformal dimension, in the case of a ground state.
Since the NS/R pairs of sectors (9)--(16) are related by simple coordinate
relabelling, we do not consider them seperately, but restrict ourselves
to (9) and~(10) from now on.

We have yet to show that our physical vertex operators form a {\it local} field
algebra. To this end, we should investigate their mutual operator products.
Equivalently, we are going to adress {\it interactions} among the $N{=}2$
string physical states ({\it cf} ref.~\cite{gom}). Generally speaking,
the required mutual locality of the vertex operators is expected to impose
some constraints on the allowed interactions.

As representatives of the ${\bf Z''}_2$-twisted $N{=}2$ string physical states,
we choose the
ground states in the sectors $\ch^{\rm NS}_{++++}$,  $\ch^{\rm R}_{++++}$,
$\ch^{\rm NS}_{+-+-}$ and  $\ch^{\rm R}_{+-+-}$, namely
$$\eqalign{
\ch^{\rm NS}_{++++}~:~&~\quad
\F\ =\ c\,e^{-\vf^2}e^{-\vf^3}\,e^{i(k^2\cdot Z^2+k^3\cdot Z^3)}~, \cr
\ch^{\rm R}_{++++}~:~&~\quad
\U^{\ua\da}\ =\ c\,e^{-\vf^2/2}e^{-\vf^3/2}\,S^{2,\ua\da}S^{3,\ua\da}\,
e^{i(k^2\cdot Z^2+k^3\cdot Z^3)}~,\cr
\ch^{\rm NS}_{+-+-}~:~&~\quad
\X^{\ua\da}\ =\ c\,e^{\tilde{\s}/2}e^{-\vf^2}e^{-\vf^3/2}\,S^{3,\ua\da}t^3\,
e^{ik^2\cdot Z^2}~,\cr
\ch^{\rm R}_{+-+-}~:~&~\quad
\L^{\ua\da}\ =\ c\,e^{\tilde{\s}/2}e^{-\vf^2/2}e^{-\vf^3}\,S^{2,\ua\da}t^3\,
e^{ik^2\cdot Z^2}~,\cr}
\eqno(7.17)$$
with $k^2\sdot k^2{+}k^3\sdot k^3=0$ in the first case and
$k^{\ua i}=0$ or $k^{\da i}=0$ in the other three,
corresponding to the chosen helicity.

On one hand, using the OPE structure of the
bosonized fields among themselves, {\it viz.}
$$\eqalign{
e^{\tilde{\s}/2}(z)\,e^{\tilde{\s}/2}(w)\
&\sim\ (z{-}w)^{1/4}\,e^{\tilde{\s}}(w)~,\cr
e^{-\vf^i/2}(z)\,e^{-\vf^i/2}(w)\ &\sim\ (z{-}w)^{-1/4}\,e^{-\vf^i}(w)~,\cr
S^{i,\ua}(z)\,S^{i,\ua}(w)\ &\sim\ (z{-}w)^{1/4}\,e^{2\f^i}(w)~,\cr
S^{i,\ua}(z)\,S^{i,\da}(w)\ &\sim\ (z{-}w)^{-1/4}~,\cr}\eqno(7.18)$$
we find that interactions between the sectors
$\ch^{\rm NS}_{++++}$ and $\ch^{\rm NS,R}_{+-+-}$ seem to be forbidden,
since the relevant OPE is {\it not local}:
$$e^{-\vf^2-\vf^3}e^{ik^2\cdot Z^2}(z)\,\,
e^{-\vf^2}e^{-\vf^3/2}e^{ik'^2\cdot Z^2}(w)\
\sim\ (z{-}w)^{-{\bf 3/2}}\,(z{-}w)^{k^2\cdot k'^2}~,\eqno(7.19)$$
and quite similarly in the other cases.~\footnote{
For consistency in the twisted string, we must put $k^3=0$ for its
untwisted states as well, effectively reducing the theory to
$(1{+}1)$ dimensions. See also ref.~\cite{gom}.}

On the other hand, each of the three triples $(\F,\U^\ua,\U^\da)$,
$(\U^\ua,\X^\da,\L^\da)$ and $(\U^\da,\X^\ua,\L^\ua)$ have {\it local} OPEs
among themselves, as can easily be checked using eq.~(7.18) again.
The non-localities do not yet mean that interactions between the triplets
are impossible, since for closed strings the left-moving (chiral) fields
still have to be combined with the right-moving ones to complete the full
vertex operators. As the example of the non-supersymmetric $O(16)\otimes O(16)$
string showed~\cite{nsu,dixhar,amp}, square root singularities in the OPEs
may disappear when the proper GSO projection is applied for modular invariance.
Hence, an {\it asymmetric} (with respect to the left- and right-moving degrees
of freedom) GSO projection may allow us to keep more than two interacting
`spacetime' fermions in the theory.
To settle this question, one should study the `bosonized lattice' of
ref.~\cite{k-w}, which in our case is a direct product of two
$(1,1)$-dimensional (half-integral) lorentzian lattices~\footnote{
The signature arises from the sign difference between the OPEs of eqs.~(4.2)
and (4.10). The twist fields are irrelevant here, since the combination
$t^3e^{\tilde{\s}/2}$ is local with any vertex operator as long as $k^3{=}0$.}
for the right-movers, and once more for the left-movers.
At the same time it is rather clear that the continuous spectral flow family
$\n\in[-\ha,\ha]$ has little chances to survive the final locality test,
except for $\n{=}0$ and $\n{=}\ha$, i.e. in the well-known NS and R~sectors.

In order to test {\it full} locality, we need to look at (tree-level)
amplitudes, i.e. correlation functions of vertex operators.
The non-vanishing 3-point amplitude for the ground state physical `scalar' in
the  $\ch^{\rm NS}_{++++}$ sector was constructed by Ooguri and
Vafa~\cite{ov}.
Admitting also the $\ch^{\rm R}_{++++}$ sector, one easily computes,
for example,
$$\VEV{\F_x\,\U^\ua_y\,\U^\da_z}\ =\ 1~,\eqno(7.20)$$
as expected from conformal invariance.
For the one boson-two fermion amplitude we also find
$$\VEV{\U^\ua_x\,\X^\da_y\,\L^\da_z}\ =\
\VEV{\U^\da_x\,\X^\ua_y\,\L^\ua_z}\ =\ 1~,\eqno(7.21)$$
which is encouraging. Ghost-number conservation does not permit other
three-point functions.

The vanishing of the tree-level bosonic 4-point
function $\VEV{\F_x\F_y\F_z\F_w}$~\cite{ov} is most easily verified in our
approach by computing the equivalent (spectral flow!) amplitude
$\VEV{\U_x\U_y\U_z\U_w}$ as follows ($\{x_i|i{=}1,\ldots,4\}=\{x,y,z,w\}$)
$$\eqalign{ \frac13 & \left[
\VEV{\U^\ua_x \U^\ua_y\U^\da_z\U^\da_w}+
\VEV{\U^\ua_x\U^\da_y\U^\ua_z\U^\da_w}+
\VEV{\U^\ua_x\U^\da_y\U^\da_z\U^\ua_w} \right] \cr
&\qquad \sim\ \VEV{\prod_i e^{-\vf/2}(x_i)}^2\,
\left[\VEV{S^\ua_x S^\ua_y S^\da_z S^\da_w}^2 +
\VEV{S^\ua_x S^\da_y S^\ua_z S^\da_w}^2 +
\VEV{S^\ua_x S^\da_y S^\da_z S^\ua_w}^2 \right]\cr
&\qquad =\ \prod_{i<j}x_{ij}^{-1}\;\Bigl[
(x{-}y)(z{-}w)-(x{-}z)(y{-}w)+(x{-}w)(y{-}z)\Bigr] =0~,\cr}\eqno(7.22)$$
where we defined $x_{ij}\equiv x_j{-}x_j$, as usual.
The crucial relative signs emerge from carefully taking into account the
suppressed cocycle operators~\cite{k-w}.
\vglue.2in

\section{Conclusions}

Our motivation for twisting the $N{=}2$ string was driven by the search for
more physical states in $N{=}2$ string theory defined in $2{+}2$ dimensions.
An arbitrary twisting implies a locally flat background~$\cb$ for $N{=}2$
string propagation, which is not just ${\bf C}^{1,1}$ but has non-trivial
global topology. Backgrounds induced by twisting generically take the form
$\cb={\bf C}^{1,1}/\G_0$, where a discrete group $\G_0$ is generated by
elements of the {\it global} symmetry group which act on the bosonic
coordinate fields $Z^{\m i}$. For flat backgrounds, the global homogeneous
symmetry group with non-trivial $Z$~action is isomorphic to
$U(1,1)\otimes{\bf Z''}_2$. Its action on the fermionic coordinates is
identical. On top of this, there exists the spectral flow labelling the
overall mismatch of bosonic and fermionic boundary conditions.
It generates a $U(1)$~family of sectors interpolating
between the NS- and R-like sectors of the $N{=}2$ string. We have found four
types of different ${\bf Z}_2$ twists leading to sixteen different sectors to
consider. One of the twists originated from the spectral flow and does not
alter the background, while two more are contained in the global
$U(1,1)$ symmetry of the $N{=}2$ string.

The quantized critical $N{=}2$ string theory can be conveniently described in
the $N{=}2$ superconformal gauge by introducing ghosts for the gauge-fixed
local symmetries and the corresponding BRST charge.
Using bosonization techniques, we constructed the spin and twist fields
which actually implement all the ${\bf Z}_2$ twists mentioned above.
Next, we identified the ground states of the sixteen ${\bf Z}_2$-twisted
sectors of the $N{=}2$ string, as well as their conformal dimensions and
local $U(1)$ charges. The `spacetime' interpretation of the twists is based on
the orbifolds and semi-spaces in eq.~(4.24) as possible target spaces for
$N{=}2$ string propagation. Half of the ground states were recognized as
candidates for `spacetime' bosons, whereas the ${\bf Z''}_2$-twisted
half suggested an interpretation as `spacetime' fermions. Not all of these
states are, however, physical states identified as representatives of
BRST cohomology classes.

The most general homogeneous monodromies of the $N{=}2$ string were shown to
be classified by the conjugacy classes of $U(1,1)$ and the ${\bf Z''}_2$ twist
of complex conjugation. They naturally split into three different groups:
elliptic, parabolic and hyperbolic, only the first type having been
considered in the past. We proposed the formula of eq.~(6.30) for the vacuum
energy of the arbitrarily $U(1,1)$-twisted ground state, by generalizing the
result known for the elliptic case. Searching for {\it massless}
ground states, we were able to restrict them to the
(possibly ${\bf Z''}_2$-twisted) identity and parabolic conjugacy classes.
The latter gives a previously unknown massless background with interesting
properties. The former leads back to ${\bf C}^{1,1}$ which yields four of
the sixteen ${\bf Z}_2$~sectors upon twisting by ${\bf Z''}_2$ and/or
spectral flow.

The presence of the spectral flow relating the R- and NS-type sectors of the
$N{=}2$ string does not mean that these are to be identified.
The spectral flow modifies fermionic boundary conditions for each
homology cycle on the world-sheet separately.
For a tree-level amplitude, this leaves {\it one} combination of the
external fermionic boundary conditions unchanged.
More specifically, we can always choose the fermionic monodromy of one
external state to be invariant under the spectral flow.
In particular, we have a $U(1)$ family of
two-point functions, indexed by the {\it relative} boson/fermion monodromy
mismatch, most of which have to vanish.
Hence, a rigid $U(1)$ label should be attached to any state. In this paper
we have only considered the ${\bf Z}_2$ subset denoted by NS and~R.

{}From the world-sheet point of view, the twistings considered in this paper
have drastic consequences for the $N{=}2$ supergravity fields as well.
We are used to the fact that gravitini may be anti-periodic around world-sheet
cycles, but e.g. the Mathur-Mukhi twist also implies a {\it double-valued\/}
abelian gauge field (see eq.~(3.7)). In other words, we are dealing with a
double cover of the world sheet. Alternatively, one may put
$A_\alpha\equiv 0$, which amounts to consistently truncating to $N{=}1$
supergravity. It can be shown \cite{2pre} that there actually exist only
{\it two\/} different GSO projections of the $N{=}2$ string, one corresponding
to the untwisted (`spacetime' bosonic) theory, the other leading to the
Mathur-Mukhi-twisted theory containing `spacetime' bosons {\it and\/} fermions.

The crucial check of our ${\bf Z}_2$-twisted ground states as candidate
physical states comes from the analysis of the BRST cohomology. We have found
that only some of these states are physical, namely one scalar, one vector,
and two spinors. Moreover, the interactions among the would-be physical
states were analyzed from the viewpoint of world-sheet locality, needed for an
unambiguous definition of the conformal field theory correlation functions.
It turned out that either three bosons or, else, one boson and two fermions may
coexist. The consequences can be found in ref.~\cite{2pre}.

Still, there is room for $N{=}2$ `spacetime' supersymmetry to be present,
although the numbers of {\it physical} boson and fermion degrees of freedom we
obtained are not enough to support the {\it maximal} `spacetime' supersymmetry
advocated in ref.~\cite{siegel}. The BRST approach to gauge
theories is well-known to be based on canonical (unitary) quantization.
The notion of unitarity is, however, quite formal in $2{+}2$ dimensions.
This is
related to the fact that in the covariant Lagrangian description of self-dual
(supersymmetric) field theories in $2{+}2$ dimensions one half of the fields
are usually the (propagating) Lagrange multipliers for the other half, and vice
versa~\cite{kng}. The BRST cohomology may have selected those half of the
(super) self-dual states with {\it positive} norms. The remaining half
represents an equal number of ghost states having {\it negative} norms, which
are, however, needed for the covariant (with respect to $SO(2,2)$) description
of $N{=}2$ string theory. This effectively doubles the number of states in
the game, and may open a way for the $N{=}4$ `spacetime' supersymmetry
of the (ghost-) extended theory. More studies are needed to resolve this
issue.
\vglue.2in

\noindent {\bf Acknowledgements}

It is a pleasure to thank J.~Bischoff, B.~Dolan, S.J.~Gates~Jr.,
C.~Preitschopf, A.~Shapere, S.~Theisen and A.A.~Tseytlin
for stimulating discussions, and the referee for constructive criticism.

\newpage

\noindent{\Large\bf Appendix A: local symmetries of the BS action}

In this Appendix, we list the infinitesimal field transformation
laws corresponding to the local two-dimensional symmetries of the
Brink-Schwarz (BS) action (3.1).

\noindent
(i) {\it Reparametrization invariance}:
$$\eqalign{
\d e^a_{\a}= & \x^{\b}\pa_{\b}e^a_{\a} + e^a_{\b}\pa_{\a}\x^{\b}~,\cr
\d \c_{\a}= & \x^{\b}\pa_{\b}\c_{\a} + \c_{\b}\pa_{\a}\x^{\b}~,\cr
\d A_{\a}= & \x^{\b}\pa_{\b}A_{\a} + A_{\b}\pa_{\a}\x^{\b}~,\cr
\d Z= & \x^{\a}\pa_{\a}Z~,~~ \d\J =\x^{\a}\pa_{\a}\J~;\cr}\eqno(A.1)$$
(ii) {\it Lorentz invariance}:
$$\eqalign{
\d e^a_{\a}= & l\ve\ud{a}{b}e^b_{\a}~,\cr
\d \c_{\a}= & -\ha l\r_3\c_{\a}~,\cr
\d A_{\a}= & \d Z = \d \J =0~;\cr}\eqno(A.2)$$
(iii) $N{=}2$ {\it extended supersymmetry}:
$$\eqalign{
\d e^a_{\a}= & -2i\bar{\ve}\r^a\c_{\a}+ {\rm h.c.}~,\cr
\d\c_{\a}= & \left( \pa_{\a}+\ha \o_{\a}\r_3-iA_{\a}\right)\ve~,\cr
\d A_{\a}= & \ve^{\b\g}\bar{\ve}\r_3\r_{\a}\left( \pa_{\b}+\ha \o_{\b}\r_3
-iA_{\b}\right)\c_{\g}+ {\rm h.c.}~,\cr
\d Z = & -2\bar{\ve}\J~,\cr
\d \J = & i\r^{\b}\ve\left( \pa_{\b}Z + 2\bar{\c}_{\b}\J\right)~;\cr}
\eqno(A.3)$$
(iv) {\it phase and chiral gauge invariances}:
$$\eqalign{
\d e^a_{\a}= & \d Z =0~,\cr
\d \c_{\a}= & i\a\c_{\a}-i\r_3\hat{\a}\c_{\a}~,\cr
\d A_{\a}= & \pa_{\a}\a +\ve\du{\a}{\b}\pa_{\b}\hat{\a}~,\cr
\d\J = & i\a\J + i\hat{\a}\r_3\J~;\cr}\eqno(A.4)$$
(v) {\it Weyl and super-Weyl invariances}:
$$\eqalign{
\d e^a_{\a}= & \s e^a_{\a}~,\cr
\d \c_{\a}= & \ha\s\c_{\a} +\r_{\a}\h~,\cr
\d A_{\a}= & \bar{\c}_{\b}\r_{\a}\r^{\b}\h+ {\rm h.c.}~,\cr
\d Z = &  0~,~~\d \J = -\ha\s\J~;\cr}\eqno(A.5)$$
where $\x$, $l$, $\ve$, $\a$, $\hat{\a}$, $\s$ and $\h$ are parameters of
reparametrization, Lorentz, $N{=}2$ extended supersymmetry, phase, chiral, Weyl
and $N{=}2$ super-Weyl (superconformal) local transformations, respectively. We
use the notation
$$ \ve^{\a\b}=e^{-1}\e^{\a\b}~,\quad \o_{\a}(e,\c)=\o_{\a}(e)+\left[
i\bar{\c}_{\b}\r^{\b}\c_{\a}+ {\rm h.c.}\right]~,\eqno(A.6)$$
where $\e^{\a\b}$ is the Levi-Civita symbol, and $\o_{\a}(e)$ is the
conventional gravitational connection in two dimensions.
\vglue.2in

\noindent{\Large\bf Appendix B: $N{=}2$ string BRST charge} \footnote{We are
grateful to Jan Bischoff for helping us to check some of the formulae listed
below.}

In this Appendix, the standard results  needed to introduce the $N{=}2$ string
BRST charge are summarized.

The OPE for the chiral parts of the matter fields representing
the $N{=}2$ string coordinates are
$$\eqalign{
Z^{i\m}(z)\ Z^{j\n}(w)\ \sim &\  - \d^{ij}\h^{\m\n}\ln (z-w)~,\cr
\j^{i\m}(z)\ \j^{j\n}(w)\ \sim &\  - \d^{ij}\h^{\m\n}\fracmm{1}{z-w}~.\cr}
\eqno(B.1)$$

The $N{=}2$ currents associated with the superconformally gauge-fixed $N{=}2$
string action take the form \footnote{Normal ordering is applied whenever
an ambiguity arises.}
$$\eqalign{T(z)\ &=\
-\ha\pa Z^+\sdot\pa Z^- +\fracm14\j^+\sdot\pa\j^- +\fracm14\j^-\sdot\pa\j^+\cr
&=\ -\ha\left(\pa Z^{i\m}\pa Z^i_{\m} - \j^{i\m}\pa\j^i_{\m}\right)
=\ \sum_{n\in{\bf Z}} L_n z^{-n-2}~,\cr}$$
$$G^\pm(z)\ =\
\pa Z^\mp\sdot\j^\pm\
=\ \pa Z^{i\m}\j^i_{\m} \pm i\ve^{ij}\pa Z^{i\m}\j^j_{\m} \
=\ \sum_{n\in{\bf Z}} G^\pm_n z^{-n-3/2}~,$$
$$J(z)\ =\
\fracm{1}{4} \j^-\sdot\j^+\
=\ \fracm{i}{4}\ve^{ij}\j^{i\m}\j^j_{\m}\
=\ \sum_{n\in{\bf Z}} J_n z^{-n-1}~,\eqno(B.2)$$
where the mode expansions on the r.h.s. of eq.~(B.2) and all the equations
below are valid for the untwisted R-type boundary conditions, for definiteness.
For the twisted and/or NS-type boundary conditions, the moding on the r.h.s.
of eq.~(B.2) and the mode expansions below have to be appropriately modified.
The currents~(B.2) form the $N{=}2$ superconformal algebra with central
extension.

The reparametrization ghosts
$$b(z)\ =\ \sum_{n\in{\bf Z}} b_n z^{-n-2}~,\qquad
c(z)\ =\ \sum_{n\in{\bf Z}} c_n z^{-n+1}~,\eqno(B.3)$$
satisfy the OPE
$$b(z)\ c(w)\ \sim\ \fracmm{1}{z-w}~,
\qquad \{c_m,b_n\}\ =\ \d_{m+n,0}~.\eqno(B.4)$$

The $N{=}2$ superconformal ghosts
$$\b^\pm(z)\ =\ \sum_{n\in{\bf Z}}\b^\pm_n z^{-n-3/2}~,\qquad
\g^\pm(z)\ =\ \sum_{n\in{\bf Z}}\g^\pm_n z^{-n+1/2}~,
\eqno(B.5)$$
satisfy the OPE
$$\b^+(z)\ \g^-(w)\ \sim\ -\fracmm{2}{z-w}~,\qquad
\b^-(z)\ \g^+(w)\ \sim\ -\fracmm{2}{z-w}~, \eqno(B.6)$$
which imply the (only nonvanishing) commutation relations
$$\[\g^+_m,\b^-_n\]\ =\ \[\g_m^-,\b_n^+\]\ =\ 2\d_{m+n,0}~.\eqno(B.7)$$

Finally, the anti-commuting $U(1)$ ghosts
$$\tilde{b}(z)\ =\ \sum_{n\in{\bf Z}}\tilde{b}_n z^{-n-1}~,\qquad
\tilde{c}(z)\ =\ \sum_{n\in{\bf Z}} \tilde{c}_n z^{-n}~,\eqno(B.8)$$
associated with the abelian local invariance of $N{=}2$ string theory,
have the OPE
$$\tilde{b}(z)\ \tilde{c}(w)\ \sim\ \fracmm{1}{z-w}~,\qquad \{\tilde{c}_m,
\tilde{b}_n\}\ =\ \d_{m+n,0}~.\eqno(B.9)$$

The full BRST-invariant generators read:
$$T_{\rm tot}=\{Q_{\rm BRST},b\}=
T-2b\pa c -(\pa b)c -\tilde{b}\pa\tilde{c}-\fracm34
[\b^-\pa\g^++\b^+\pa\g^-]-\fracm14[\g^+\pa\b^-+\g^-\pa\b^+],$$
$$G^\pm_{\rm tot}=\[Q_{\rm BRST},\b^\pm\]=
G^\pm-2b\g^\pm-4\tilde{b}\pa\g^\pm-2(\pa\tilde{b})\g^\pm+\fracm32
\b^\pm\pa c+(\pa\b^\pm)c+\ha\b^\pm\tilde{c}~,$$
$$J_{\rm tot}=\{Q_{\rm BRST},\tilde{b}\}=
J +\tilde{b}\pa c+(\pa\tilde{b})c+\fracm14[\b^+\g^--\b^-\g^+]~.\eqno(B.10)$$

They imply, in particular, the following mode expansions: \footnote{
Summation over repeated indices is always understood.}
$$\eqalign{L^{\rm tot}_m\ =&\ \{Q_{\rm BRST},b_m\}\ =\
L_m + (m-n)b_{m+n}c_{-n}-n\tilde{b}_{n+m}\tilde{c}_{-n}\cr
&+\fracm14(m-2n)[\b^-_{m+n}\g^+_{-n}+\b^+_{m+n}\g^-_{-n}] -A_0\d_{m,0}~,\cr}$$
$$\eqalign{G^{\pm,\rm tot}_m\ =&\ \[Q_{\rm BRST},\b^\pm_m\]\ =\
G^\pm_m- 2b_{m+n}\g^\pm_{-n}-2(m-n)\tilde{b}_{m+n}\g^\pm_{-n}\cr
&+(\ha n-m)\b^\pm_{m+n}c_{-n}+\ha\b^\pm_{m+n}\tilde{c}_{-n}~,\cr}$$
$$J^{\rm tot}_m=\{Q_{\rm BRST},\tilde{b}_m\}=
J_m- m\tilde{b}_{m+n}c_{-n}+\fracm14[\b^+_{m+n}\g^-_{-n}-\b^-_{m+n}\g^+_{-n}]
-B_0\d_{m,0}~,\eqno(B.11)$$
where normal-ordering ambiguity constants $A_0$ and $B_0$ have been
introduced.

The $N{=}2$ BRST charge is given by
$$Q_{\rm BRST}\ =\ \oint_0 \fracmm{dz}{2\p i}\,j_{\rm BRST}(z)~,\eqno(B.12)$$
with the BRST current having the form
$$\eqalign{j_{\rm BRST}(z)\ =&\ cT +bc\pa c+\tilde{b}c\pa\tilde{c}
-\fracm34c[\b^-\pa\g^++\b^+\pa\g^-] -\fracm14c[\g^+\pa\b^-+\g^-\pa\b^+]\cr
&+\ha[\g^-G^+ +\g^+G^-] + \tilde{c}J\cr
&-\g^-\g^+b +[\g^-\pa\g^+-\g^+\pa\g^-]\tilde{b}
+\fracm14\tilde{c}[\b^+\g^--\b^-\g^+]\cr
&+\fracm38\pa[c(\b^+\g^-+\b^-\g^+)]~.\cr}\eqno(B.13)$$
In terms of real fields,
$$\eqalign{j_{\rm BRST}(z)\ =&\ cT +bc\pa c+\tilde{b}c\pa\tilde{c}
-\fracm32c[\b^2\pa\g^2+\b^3\pa\g^3] -\fracm12c[\g^2\pa\b^2+\g^3\pa\b^3]\cr
&+\g^2[\pa Z^2\sdot\j^2+\pa Z^3\sdot\j^3]
+\g^3[\pa Z^2\sdot\j^3-\pa Z^3\sdot\j^2]+\fracm{i}2\tilde{c}\j^2\sdot\j^3\cr
&-[\g^2\g^2+\g^3\g^3]b+2i[\g^2\pa\g^3-\g^3\pa\g^2]\tilde{b}
+\fracm{i}2\tilde{c}[\g^2\b^3-\g^3\b^2]\cr
&+\fracm34\pa[c(\b^2\g^2+\b^3\g^3)]~.\cr}\eqno(B.14)$$
The total derivative terms have been adjusted to make the current
$j_{\rm BRST}$ BRST-exact and turn it into a conformal primary of
$h{=}1$ and $e{=}0$. It follows
$$\eqalign{Q_{\rm BRST}\ =&\
c_{-n}L_n +\ha\g^-_{-n}G^+_n +\ha\g^+_{-n}G^-_n+\ha\tilde{c}_{-n}J_n\cr
&-\ha(m-n)c_{-m}c_{-n}b_{m+n}-\g^-_{-m}\g^+_{-n}b_{m+n}
-(m-n)\g^-_{-m}\g^+_{-n}\tilde{b}_{m+n}\cr
&+nc_{-m}\tilde{c}_{-n}\tilde{b}_{m+n}
+\fracm14(m-2n)c_{-m}[\b^+_{m+n}\g^-_{-n}+\b^-_{m+n}\g^+_{-n}]\cr
& +\fracm14\tilde{c}_{-m}[\b^+_{m+n}\g^-_{-n}-\b^-_{m+n}\g^+_{-n}]
-A_0c_0 -B_0\tilde{c}_0~.\cr}\eqno(B.15)$$

The BRST generators (B.11) satisfy the $N{=}2$ superconformal algebra without
central extension, while the BRST charge $Q_{\rm BRST}$ is nilpotent,
$Q_{\rm BRST}^2=0$, when the complex target space dimension is two and the
intercept takes its critical value under the given boundary conditions.

\end{document}